\documentclass[12pt,a4paper]{article}
\usepackage{graphicx}
\usepackage{dcolumn}
\usepackage{bm}
\usepackage{amsmath,amsfonts,amssymb}
\usepackage{slashed}
\usepackage{braket,xcolor}
\usepackage{verbatim}
\usepackage{subcaption}
\usepackage{multirow}
\usepackage{amsfonts,mathtools,resizegather}
\usepackage[utf8]{inputenc}
\usepackage{caption,jheppub}

\title{\boldmath {Operator growth and Krylov construction in dissipative open quantum systems}
 }
\author[a]{Aranya Bhattacharya,}
\author[a,b]{Pratik Nandy,}
\author[a]{Pingal Pratyush Nath,}
\author[a]{and Himanshu Sahu}

\affiliation[a]{Centre for High Energy Physics, Indian Institute of Science,\\ C.V. Raman Avenue, Bangalore 560012, India.}
\affiliation[b]{Center for Gravitational Physics and Quantum Information,\\ Yukawa Institute for Theoretical Physics, Kyoto University,\\
Kitashirakawa Oiwakecho, Sakyo-ku, Kyoto 606-8502, Japan}

\emailAdd{aranyab@iisc.ac.in, 
pratik@yukawa.kyoto-u.ac.jp, pingalnath@iisc.ac.in, himanshusah1@iisc.ac.in}

\abstract{Inspired by the universal operator growth hypothesis, we extend the formalism of Krylov construction in dissipative open quantum systems connected to a Markovian bath. Our construction is based upon the modification of the Liouvillian superoperator by the appropriate Lindbladian, thereby following the vectorized Lanczos algorithm and the Arnoldi iteration. This is well justified due to the incorporation of non-Hermitian effects due to the environment. We study the growth of Lanczos coefficients in the transverse field Ising model (integrable and chaotic limits) for boundary amplitude damping and bulk dephasing. Although the direct implementation of the Lanczos algorithm fails to give physically meaningful results, the Arnoldi iteration retains the generic nature of the integrability and chaos as well as the signature of non-Hermiticity through separate sets of coefficients (Arnoldi coefficients) even after including the dissipative environment. Our results suggest that the Arnoldi iteration is meaningful and more appropriate in dealing with open systems.}

\footnotetext{Authors' names are listed in the alphabetical order.}

\begin{document}
\maketitle
\flushbottom
\section{Introduction}

Integrability and chaos has remained a long-standing problem in quantum many-body systems. In a closed system, the integrability is marked by a substantial number of conserved quantities \cite{PhysRevLett.111.127201, ROS2015420, PhysRevB.91.085425}. This shows up in the level statistics of the unfolded spectrum, which follow the Poisson distribution, known as the Berry-Tabor conjecture \cite{Berry_Tabor}. In contrast, Wigner-Dyson statistics is noticeable for the chaotic systems, often referred to as Bohigas-Giannoni-Schmit (BGS) conjecture \cite{Wigner, BGS}. However, dissipation and decoherence are ubiquitous in nature, making any evolution non-unitary. The effect of decoherence has been a growing interest in the research of many-body systems, and field theory \cite{PhysRevLett.118.140403, Beau:2017xaj, DelCampo:2019afl, Xu:2020wky}. Decoherence may not always be associated with the environment. It can happen solely due to the random fluctuations, noise, and random phase changes in the system \cite{braun2003dissipative, Xu:2018bhd}. Hence, it is of practical interest to extend the notion of integrability and chaos in open systems where the system interacts with the environment. Indeed, Grobe, Haake, and Sommers (GHS) generalized the above ideas to such systems \cite{PhysRevLett.61.1899}, governed by Markovian dynamics. Instead of the Hermitian Hamiltonian, one considers the level statistics of the corresponding Lindbladian supplemented by the appropriate jump operators, which capture the dissipation. The eigenvalues of the Lindbladian can be real or complex (in pairs), and one can devise an appropriate version of the level statistics. While integrable systems exhibit Poisson statistics, chaotic systems follow the distribution of Ginibre ensemble \cite{Ginibre, PhysRevLett.61.1899, PhysRevLett.123.254101, PhysRevResearch.2.023286}. Thus, level statistics provides a powerful way to detect non-integrability in open systems.

Operator growth, on the other hand, provides an indirect yet illuminating way to capture such behavior. The idea stems from the fact that under Heisenberg evolution, a simple operator becomes complicated as time progresses. A piece of local information supported on a single or few sites spread to the whole system, leading to the thermalization of chaotic systems \cite{PhysRevA.43.2046, PhysRevE.50.888, PhysRevLett.98.050405, Rigol_2008}. Depending on whether the Hamiltonian is integrable or chaotic, operators grow differently. Although such growth has been well understood in terms of the out-of-time-ordered functions (OTOC), it has recently been revived by the universal operator growth hypothesis (UOGH) \cite{Parker:2018yvk}, which extends the operator growth beyond the semi-classical limit. It states that the chaotic systems show the fastest growth of Lanczos coefficients (defined later), although all the systems with the fastest growth may not be chaotic \cite{Dymarsky:2021bjq, Bhattacharjee:2022vlt}. Recently, the UOGH has gained much interest from the many-body perspective, conformal field theory and holography \cite{Avdoshkin:2019trj, Dymarsky:2019elm, Barbon:2019wsy, Rabinovici:2020ryf, Jian:2020qpp, PhysRevE.104.034112, Cao:2020zls, Magan:2020iac, Rabinovici:2021qqt, Yates:2021asz, PhysRevLett.124.206803, PhysRevB.102.195419, Kim:2021okd, Caputa:2021ori, Trigueros:2021rwj,  Bhattacharjee:2022vlt, Caputa:2021sib, Patramanis:2021lkx, Hornedal:2022pkc, Heveling:2022hth, Kar:2021nbm, Muck:2022xfc, Banerjee:2022ime, Heveling:2022orr}.

In this work, we take a first step to extend the operator growth and the underlying formalism into dissipative open quantum systems. Our setup is the well-known transverse field Ising model (TFIM) interacting with the environment. In order to see the growth, one needs to expand the operator in terms of a Krylov basis. The standard approach is to apply the well-known Lanczos algorithm, which is efficient for the unitary evolution. This is in contrast to the non-unitary evolution of the open systems where the Lindbladian is non-Hermitian. Still, the direct application leads to some intriguing results capturing the breakdown of Hermitian methods in the case of non-Hermitian cases. For the integrable case, as soon as the system starts interacting with the environment, i.e., non-Hermiticity sets in non-trivially, the sublinear growth of the Lanczos coefficients tends to become linear. Furthermore, the growth becomes indiscernible between the integrable and chaotic limit for large Lanczos coefficients, which appears to grow forever. This behavior is clearly unphysical due to the finite-dimensional restriction of the Krylov subspace. This effect of decoherence must be distinguished from chaos \cite{Yoshida:2018vly}. Hence, we need a more refined algorithm that correctly incorporates such a non-Hermitian effect. The Arnoldi iteration, a well-known iterative algorithm to find the eigenvalues and eigenvectors of non-Hermitian matrices, achieves this. Of course, when the evolution becomes unitary, the Arnoldi iteration reduces to the usual Lanczos algorithm. In the following sections, we ask the question of whether the successful implementation of the Arnoldi iteration truly enables tracking the integrability or chaotic nature of the system. Also, to be fully sensitive to the conditions and parameters involved, we will also ask whether the Arnoldi iteration can distinguish between different values of the non-Hermitian couplings involved in the study. To our satisfaction, we find the answers to both the questions are positive.

The paper is structured as follows. In section \ref{krylov}, we briefly discuss the construction of the Krylov basis. Section \ref{open} introduces the open system, especially the model with computation of Lanczos coefficients, including the dissipative environment. We discuss the details of the Arnoldi iteration in section \ref{arnoldi}. We conclude the paper with a brief summary and outlook in section \ref{conc}.


\section{Operator growth in Krylov basis} \label{krylov}

In this section, we briefly review the basic formulation of Lanczos algorithm for the construction of orthonormal Krylov basis \cite{viswanath1994recursion, Parker:2018yvk}. Consider an initial operator $\mathcal{O}_{0}$ at $t=0$, sometimes known as the seed operator. Under the unitary time evolution governed by a time-independent Hamiltonian $H$, the time-evolved operator $\mathcal{O}(t)$ is written by the Heisenberg evolution as
\begin{equation}
    \mathcal{O}(t) = e^{i H t}\,\mathcal{O}_{0}\,e^{-i H t} = e^{i \mathcal{L} t} \mathcal{O}_{0}\,,
\end{equation}
where $\mathcal{L} \, [\bullet] = [H, \bullet]$ is the Liouvillian superoperator. The explicit time evolution can be obtained by using the Baker-Campbell-Hausdorff (BCH) formula, which involves nested commutators of the form $[H, [H,\mathcal{O}_0], \cdots]$ obtained by the repeated application Liouvillian $\mathcal{L}^n \mathcal{O}_0$. For a local seed operator $\mathcal{O}_0$, these nested commutators indicate the increasing degree of non-locality, and they eventually become extremely difficult to compute. However, one can circumvent this difficulty by constructing an orthonormal set of basis $|\mathcal{O}_n)$, with $n = 0, 1, 2, \cdots, \mathcal{K}-1$, known as Krylov basis. The dimensionality of the Krylov space $\mathcal{K}$ is constrained to $1 \leq \mathcal{K} \leq D^2 - D+1$ \cite{Rabinovici:2020ryf}, where $D$ is the dimension of the Hilbert space. This makes the problem much simpler to tackle. On such a basis, the Liouvillian takes the tridiagonal form
\begin{align}
    \mathcal{L} |\mathcal{O}_n) = b_n |\mathcal{O}_{n-1}) + b_{n+1} |\mathcal{O}_{n+1})\,, \label{tri}
\end{align}
where the coefficients $b_n$'s are known as Lanczos coefficients with $b_0 = 0$. for a finite-dimensional system with Krylov dimension $\mathcal{K}$, $b_{n}$ halts at $n=\mathcal{K}-1$ with the Krylov basis $|\mathcal{O}_\mathcal{K})$ such that $b_{\mathcal{K}} = 0$. The evolved operator $\mathcal{O}(t)$ can be expanded as 
\begin{equation}
    |\mathcal{O}(t)) = \sum_{n = 0}^{\mathcal{K}-1} i^{n}\phi_{n}(t)|\mathcal{O}_{n})\,,
\end{equation}
where the ``wavefunctions'' $\phi_{n}(t) = i^{-n}(\mathcal{O}_{n}|\mathcal{O}(t))$ mimic the probability density and satisfies the unitarity constraint \cite{Parker:2018yvk}
\begin{equation}
    \sum_{n}|\phi_{n}(t)|^{2}=1\,.
\end{equation}
The first wavefunction is known as the auto-correlation function and is given by
\begin{equation}
      \phi_0 = (\mathcal{O}_{0}|\mathcal{O}(t))\,.
\end{equation}
It is worth mentioning here that the auto-correlation function can also produce the exact same set of $b_n$'s by the method of moments with the corresponding recursion relation \cite{Parker:2018yvk}. At this stage, one should properly define the inner product $(A|B)$ of two operators $A$ and $B$. One usually considers the infinite-temperature Wightmann inner product given by $(A|B) =  \frac{1}{D} \mathrm{Tr}(A^{\dagger}B)$. It is to be noted that the notion of the inner product is not unique\footnote{An alternate choice of the inner product could be with respect to the non-equilibrium steady state density matrix $\rho_{\mathrm{NESS}}$, so that $\mathcal{L}_S \{\rho_{\mathrm{NESS}}\}=0$. This is under the assumption that the Lindbladian has at least one stationary state. Hence, the corresponding definition of inner product will be $(A|B)=\frac{1}{D}\mathrm{Tr}(\rho_{\mathrm{NESS}} A^{\dagger}B)$. On the other hand, the inner product with equilibrium density matrix $\rho_{\infty} = \lim_{t \rightarrow \infty} \rho(t)$ is generically given by $(A|B) \propto \mathrm{Tr}(\rho_{\infty} A^{\dagger}B)$, with $\rho_{\infty} \propto I$, when the jump operators are Hermitian \cite{PhysRevB.98.020202}. Nevertheless, in this paper, we stick to the choice of the equilibrium infinite temperature inner product. We thank Xiangyu Cao for pointing this out.}, and one can, in principle, define another set of the inner product.  However, this definition is a standard choice in the literature \cite{Parker:2018yvk}.

The wavefunctions are recursively related by the Lanczos coefficients by a recursive differential equation of the form
\begin{equation}\label{diffeq}
    \dot{\phi}_n(t)= b_n \phi_{n-1}(t)-b_{n+1}\phi_{n+1}(t).
\end{equation}
with $\phi_{-1} (t) = 0$ and $\phi_n (0) = \delta_{n 0}$. This recursion can be visualized as a particle hopping in a lattice where the hopping amplitudes are encoded in the Lanczos coefficients. The average expectation of the particle hopping in the lattice site is defined as the Krylov complexity $ K(t) = \sum_{n}n |\phi_{n}(t)|^{2}$, and they contain exactly the same information as the Lanczos coefficients \cite{Parker:2018yvk, Barbon:2019wsy}. For our purpose, we see that Lanczos coefficients suffice to reach the main conclusion, without the need of computing the Krylov complexity.


\section{Open systems: The Lindbladian} \label{open}

In open systems, i.e., systems that interact with the environment, the evolution of operators is subtle. The total Hamiltonian can be separated into three parts
\begin{equation}
    H_{\mathrm{SE}}=H_{\mathrm{S}} \otimes I_{\mathrm{E}} + I_{\mathrm{S}} \otimes H_{\mathrm{E}} +  H_I\,,
\end{equation}
where the full dynamics is governed by the system Hamiltonian $H_{\mathrm{S}}$ together with the environment Hamiltonian $H_{\mathrm{E}}$. Here $H_I$ denotes the interaction between the system and the environment \cite{Breuer2007}, and $I_{\mathrm{S}}$ and $I_{\mathrm{E}}$ denotes the identity operators acting on the system and the environment respectively. A simple way to write the interacting Hamiltonian is the following
\begin{equation}
    H_I=\sum_i \alpha_i \, S_i \otimes E_i\,,
\end{equation}
where $S_i$ and $E_i$ are operators in the system, and environment Hilbert space, respectively. $\alpha_i$ denotes the coupling associated with this interaction. The open systems are important physically as well as experimentally since the preparation of an ideal isolated system is almost impossible. Hence, most of the experimental setups are essentially open systems with always some interaction with the environment. The dynamics of the entire system is given by the von Neumann equation $\dot{\rho}_{\mathrm{SE}}(t)=-i[H_{\mathrm{SE}},\rho_{\mathrm{SE}}(t)]$, where $\rho_{\mathrm{SE}}$ is the density matrix of the total system. This can also be understood as the equivalent to the Schr\"odinger equation for the pure state of the full system. The well-known Heisenberg equation of motion of an operator $O_{\mathrm{SE}}(0)$ reads
\begin{equation}
	    \dot{O}_{\mathrm{SE}}(t) =i [H_{\mathrm{SE}}, O_{\mathrm{SE}}(t)]=i \mathcal{L}_c O_{\mathrm{SE}}(t) \,,
\end{equation}
where $\mathcal{L}_c \,\bullet =[H_{\mathrm{SE}},\,\bullet]$ is known as the Liouvillian for the closed system. We are especially interested in studying the dynamics of the system which lives in the Hilbert space $\mathcal{H}_{\mathrm{S}}$. For that, we will consider the reduced density matrix $\rho_{\mathrm{S}}(t)=\mathrm{Tr}_{\mathrm{E}}[\rho_{\mathrm{SE}}(t)]$ by tracing out the degrees of freedom from the environment. The system density matrix evolves as $\rho_{\mathrm{S}}(t) = \sum_j E_j \rho_{\mathrm{S}}(0) E_j^{\dagger}$, where $\rho_{\mathrm{S}}(0)$ is the system density matrix at $t=0$, and $E_j$'s are known as the Kraus operators satisfying the normalization $\sum_j E_j^{\dagger} E_j = 1$. They incorporate all the information of the environment and have imprints left of their interaction with its environment. Since we are interested in the system dynamics, we want to study operator dynamics in the system part eventually. In other words, our knowledge of the environment is limited to its interaction with the environment through system parameters only. This is expected since, it is impossible to know the detailed dynamics of the full universe. 

\begin{figure}[t]
   \centering
\includegraphics[width=0.70\textwidth]{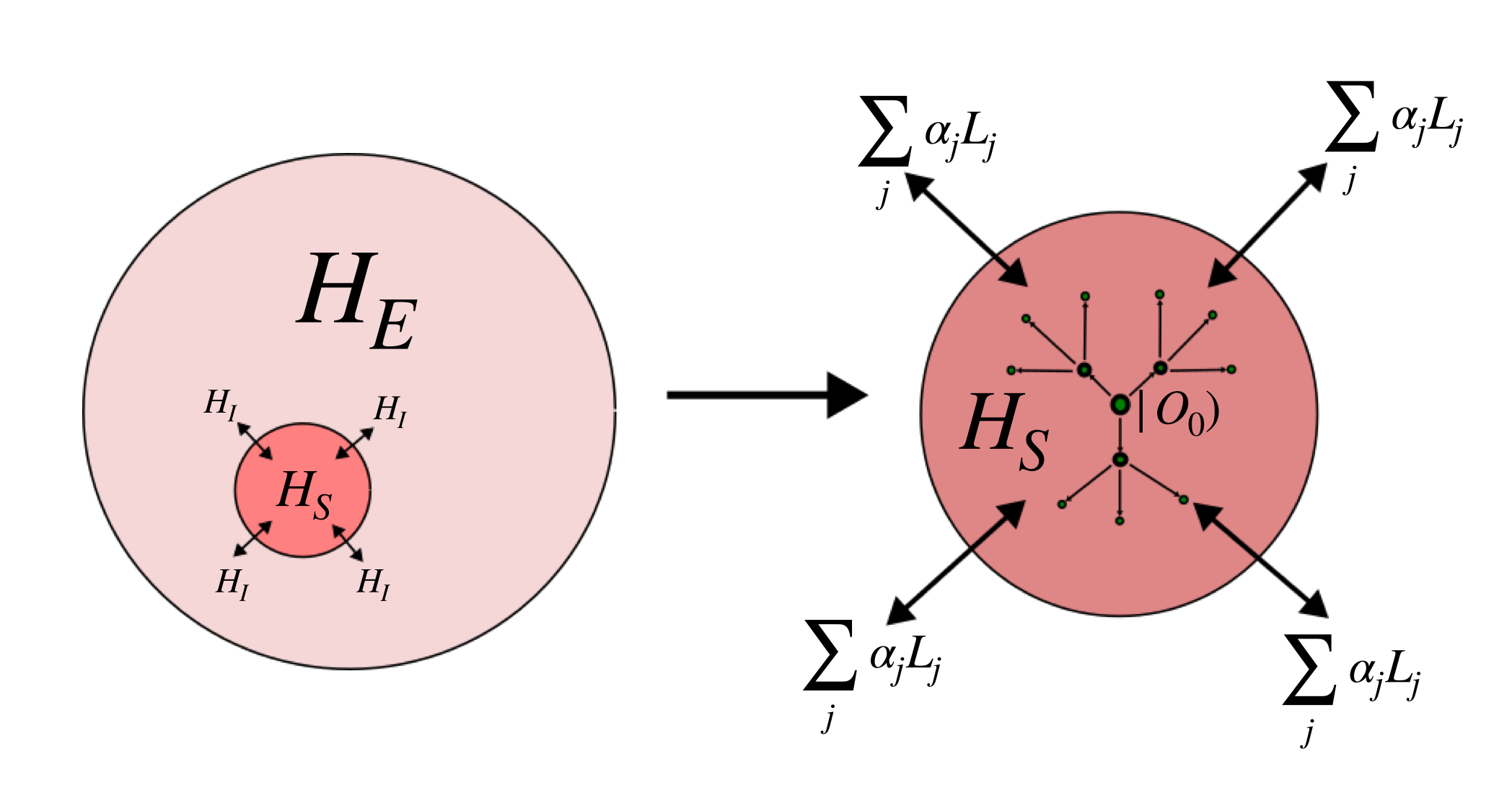}
\caption{A schematic diagram for the open system dynamics. In the L.H.S, we have the system plus the environment. R.H.S shows the system dynamics only, with operator growing within system Hilbert space under the action of system Hamiltonian $H_\mathrm{S}$ and the Lindblad operators $\sum_j \alpha_j L_j$, encoding the interaction with the environment.} \label{fig:opendiagram}
\end{figure}

We are typically interested in the Markovian dynamics which essentially leads to a small time expansion of the form 
\begin{equation}
    \rho_{\mathrm{S}}(t+dt) = \rho_{\mathrm{S}}(t) + \mathcal{O}(dt)\,.
\end{equation}
It means that the system density matrix $\rho_{\mathrm{S}}(t+dt)$ at a time $t+dt$ remembers its past history only infinitesimally up to $\rho_{\mathrm{S}}(t)$. Going forward, we drop the index ``S'' from the density matrix and it will always imply the system density matrix unless specified otherwise. The Born-Markovian dynamics of the system density matrix is given by the Lindblad master equation \cite{Lindblad1976, Gorini}
\begin{align}
	\dot{\rho} =-i[H,\rho]+\sum_k \big[L_k \rho L_k^{\dagger}-\frac{1}{2}\{L_k^{\dagger} L_k,\rho\} \big]\,,
\end{align}
or compactly as $\dot{\rho}(t) =-i \mathcal{L}_o \rho(t)$, where $\mathcal{L}_o$ is known as the Lindbladian superoperator, $H$ is the Hamiltonian of the system, and $\{\,,\,\}$ is the usual anti-commutator. This can be thought as the generator of a dynamical map $V(t) = e^{i \mathcal{L}_o t}$ on the space of the reduced density matrix of the system $V(t) : \mathcal{S}(H_{\mathrm{S}}) \rightarrow \mathcal{S}(H_{\mathrm{S}})$ and satisfying the semi-group property  \cite{Breuer2007}. The operators $\{L_j\}$ are known as the Lindblad or the jump operators which are constructed from the Kraus operators $E_j$ in the interaction part of the Hamiltonian $H_I$ along with a coupling parameter $\alpha$. \par

The Schr\"odinger evolution for the density matrix under the full Hamiltonian $H_{\mathrm{SE}}$ is very similar to the Heisenberg evolution of an operator. Hence, a system operator $O \equiv O(0)$ evolves as $\dot{O} (t) = i \mathcal{L}_o O(t)$, where the Lindbladian is \cite{Breuer2007, lidar2019lecture}
\begin{align}
	\label{Lindbladian Heisenberg}
	\mathcal{L}_o [\bullet] = [H,\bullet]-i\sum_{k}\big[L_k^{\dagger} \bullet L_k-\frac{1}{2}\{L_k^{\dagger}L_k,\bullet\}\big]\,,
\end{align}
where the subscript ``o''  indicates that the system is open. Here, as before, $H$ is the system Hamiltonian, thus including the Hermitian part only, and the $\bullet$ can be replaced by any generic operator $O(t)$. Note a significant difference in the Lindbladian from the Liouvillian. The Lindbladian has a non-Hermitian part apart from the usual Hermitian commutator. This results in a non-unitary dynamics of the system state density matrix $\rho$ as well as the system operator $O(t)$. This is a consequence of dissipation coming in due to interaction with the environment. In other words, the smaller the coupling parameters in the non-Hermitian part are, the closer the system is to an isolated system. The dynamics is summarized pictorially in Fig.\,\ref{fig:opendiagram}.

\subsection{The Hamiltonian: Transverse field Ising model}
We consider the paradigmatic example of the $1d$ transverse-field Ising model (TFIM) given by the Hamiltonian with open boundary condition \cite{PhysRevB.82.174411}
\begin{align}
    H_{\mathrm{TFIM}} = - \sum_{j=1}^{N-1}  \sigma^{z}_{j} \sigma^{z}_{j+1} - g \sum_{j=1}^{N} \sigma^{x}_j - h \sum_{j=1}^{N} \sigma^{z}_{j}\,, \label{tfim}
\end{align}
where $g$ and $h$ are the couplings. The above Hamiltonian is integrable for all $g$ and $h=0$, where it can be mapped to the free-fermionic model by Jordan-Wigner transformation \cite{sachdev_2011}. On the other hand, it is non-integrable for nonzero $g, h$. For the integrable case, we choose  $g=1, h=0$, and for the chaotic case we choose  $g=-1.05, h=0.5$, for which this Hamiltonian is strongly
chaotic \cite{PhysRevLett.106.050405, Roberts:2014isa}.
	
The interaction between the system and the environment is encoded in the jump operators. We consider the following jump operators \cite{PhysRevLett.123.254101}
\begin{align}
	L_{-1} &= \sqrt{\alpha}\,\sigma_1^{+}\,,~~~~\, L_0=\sqrt{\alpha}\,\sigma_1^{-}\,, \nonumber \\
	L_{N+1} & = \sqrt{\alpha}\,\sigma_N^{+}\,, ~~~~L_{N+2} = \sqrt{\alpha}\,\sigma_N^{-}\,, 
	~~~~L_i =\sqrt{\gamma} \,\sigma_{i}^z\,, ~~~~ i =1, 2, \cdots, N\,.\label{jump}
\end{align}
where $\sigma_j^{\pm}= (\sigma_j^{x}\pm i \sigma_j^y)/2$. The operators $L_k$ with $k = -1,\, 0,\, N+1,\, \text{and}\, N+2$ encode the boundary amplitude damping with amplitude $\alpha > 0$ while the operators $L_i, ~i =1, 2, \cdots, N$ denote the bulk dephasing with amplitude $\gamma > 0$. However, other choices with various degrees of magnitude are also possible and some of choices are described in the Appendix \ref{appa}.
	
Note that here we are using open boundary condition and we are indifferent towards the specific nature of the environment. However, through our choice of $\{L_j\}$ operators, we assume that the significant interaction with the environment is not only present at the boundary lattice points ($1^{\mathrm{st}}$ and $N^{\mathrm{th}}$), but they also affect the bulk sites of the system. We are interested in studying the operator dynamics of some system operator $O$ of the open system under Heisenberg evolution. We first apply the usual Lanczos algorithm and subsequently the Arnoldi iteration to the operator growth in the above system.

\subsection{Lanczos algorithm with Lindbladian}
We apply the Lanczos algorithm \cite{viswanath1994recursion, Parker:2018yvk} to the TFIM with the Liouvillian $\mathcal{L}_c$ replaced by the Lindbladian $\mathcal{L}_o$. The significant change here is that the Lindbladian is non-Hermitian for $\alpha, \gamma \neq0$. For $\alpha = \gamma =0$, it boils down to the usual definition of Liouvillian. However, the Heisenberg equation of motion remains the same in terms of Lindbladian and reads
\begin{align}
    O(t) =e^{i\mathcal{L}_o t} O\,.
\end{align}

\begin{figure}[t]
   \centering
\begin{subfigure}[b]{0.46\textwidth}
\centering
\includegraphics[width=\textwidth]{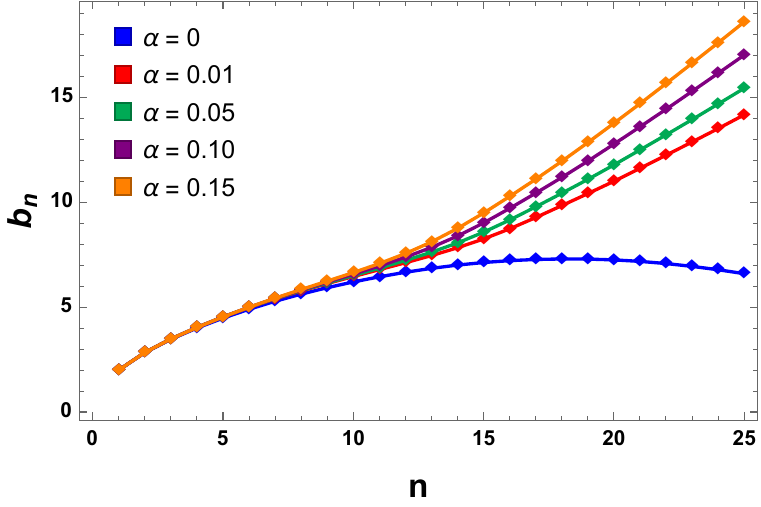}
\caption{}
\end{subfigure}
\hfill
\begin{subfigure}[b]{0.46\textwidth}
\centering
\includegraphics[width=\textwidth]{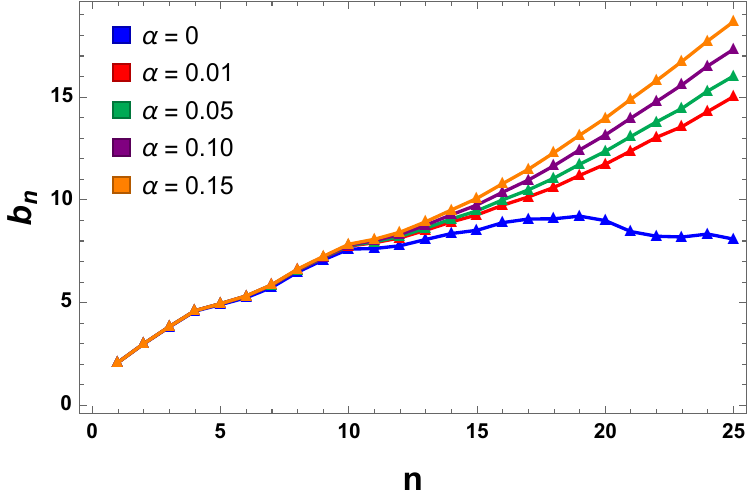}
\caption{}
\end{subfigure}
\hfill
\begin{subfigure}[b]{0.46\textwidth}
\centering
\includegraphics[width=\textwidth]{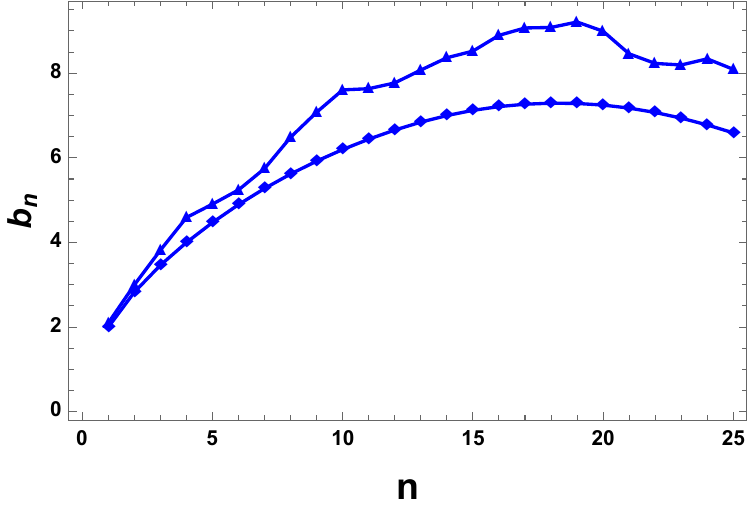}
\caption{}
\end{subfigure}
\hfill
\begin{subfigure}[b]{0.46\textwidth}
\centering
\includegraphics[width=\textwidth]{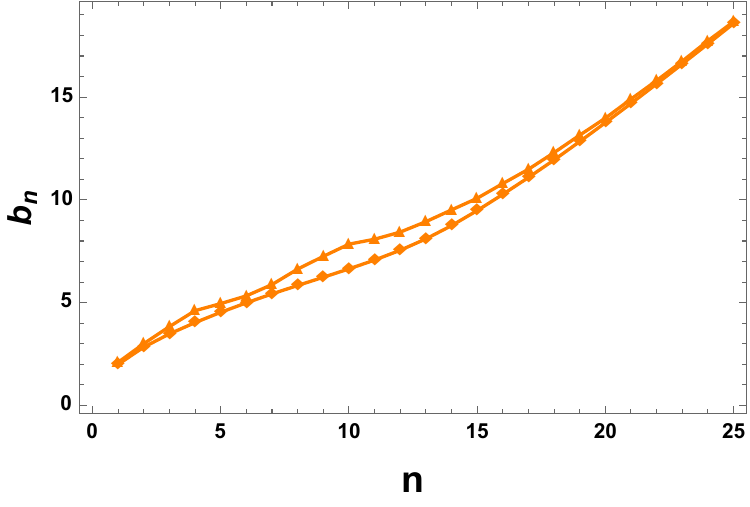}
\caption{}
\end{subfigure}
\caption{Growth of $b_n$'s of the operator $Z_3$ in (a) integrable and (b) chaotic limits by implementing the Lanczos algorithm. We individually compare the growth for (c) $\alpha =0$ (d) $\alpha = 0.15$ in both integrable (marked with diamond) and chaotic limits (marked with triangle). We choose lattice size $N=6$ and the bulk dephasing (on all sites) parameter is chosen as $\gamma = 0.1$. The blue curves are for closed system i.e., for $\alpha = 0$ and $\gamma = 0$.} \label{fig:Lan1}
\end{figure}

We apply the algorithm to the ``vectorized'' form of the Lindbladian. In terms of the density matrix, $\rho=\sum_{i} p_i |\psi_i\rangle\langle\psi_i|$, this vectorization implies transforming the  $(2^N \times 2^N)$ square matrix to a $(4^N \times 1)$ column matrix equivalent to a state, a form of operator-state mapping, usually known as ``Choi–Jamiołkowski isomorphism'' \cite{CHOI1975285, JAMIOLKOWSKI1972275} or channel-state duality \cite{PhysRevA.87.022310}. The Lindbladian also changes from a $(2^N \times 2^N)$ matrix to a $(4^N \times 4^N)$ matrix superoperator acting on the state version of the operator, in the doubled Hilbert space. In such enlarged space, the Lindbladian is given by \cite{PhysRevX.10.021019, vec}
\begin{align}
    \mathcal{L}_o = (I \otimes H - H^T \otimes I) + \frac{i}{2} \sum_k \bigg(I \otimes L_k^{\dagger} L_k + L_k^T L_k^{*} \otimes I - 2 \, L_k^{T} \otimes L_k^{\dagger} \bigg)\,,
\end{align}
where $k \in \{-1, 0, 1, \cdots, N, N+1, N+2\}$, and the jump operators are given by Eq.\eqref{jump}. Here the ``$T$'' and ``$*$'' implies the transposition and the conjugation operation, respectively. We choose the initial operator located at site three ($Z_3$) in a system size $N=6$. The jump operators \eqref{jump} acts on the boundary as well as the bulk sites (for other choices depending upon different coupling choices, see Appendix \ref{appa}). The growth of the Lanczos coefficients is shown in Fig.\,\ref{fig:Lan1}. The dissipation parameters are chosen as $\alpha = 0.01, 0.05, 0.1, 0.15$ and $\gamma = 0.1$.

We see that for $\alpha = 0,\, \gamma=0$, the Lanczos coefficients follow the sublinear growth in the integrable limit while they show linear growth in the chaotic limit, as one expects for a closed system (see Fig.\,\ref{fig:Lan1}(c)). However, as the non-Hermitian coupling parameters $\alpha$ and $\gamma$ increase, the plots deviate quickly from the Hermitian counterpart. This happens even before the finite-size effects start kicking in. For large $n$, the growth of the Lanczos coefficients for both open and closed systems appears to grow in a similar fashion. This effect solely happens due to decoherence, not due to chaos (see closely related conclusions in \cite{Xu:2018bhd, Xu:2020wky}). In other words, once the environmental effects set in, the open and closed systems become barely discernible under a Hermitian recursive algorithm. This is because the non-Hermitian terms of the Lindbladian represent dissipation in the system. This dissipation can change the system's behavior in such a way that the integrable and chaotic regimes become indistinguishable in terms of the behavior of the growth of the Lanczos coefficients. This is consistent with the behaviors in OTOC as found in \cite{PhysRevA.103.062214, Andreadakis:2022hqt}.  Moreover, the above conclusion shows a clear breakdown of the Hermitian Lanczos algorithm in the case of non-Hermitian systems. It also appears that the growth overwhelms the finite-size corrections, and it keeps growing for any non-zero coupling. This does not appear to be physically sensible as one expects these coefficients to eventually come down to zero at the end of the Krylov space. Hence the systematic growth of support for the operator loses its meaning when we look at the Lanczos coefficients. 

Another reflection of this breakdown of the Hermitian method is the bad orthonormality between the Krylov basis vectors for the open system when the Hermitian Lanczos algorithm is implemented. The orthonormality, in this case, is only up to $10^{-1}$, i.e., the vectors which are supposed to be orthonormal to each other seem to have an overlap in the order of $10^{-2}$. Here we want to treat the non-Hermitian system as a slightly perturbed system from the Hermitian (closed) one. Hence, we implemented the Lanczos algorithm that is specific for closed systems, where there are no diagonal terms assumed in the tridiagonal form of the Lindbladian, which is usually the case for a Hermitian Liouvillian. However, one can also consider the more generalized version of the Lanczos algorithm, where the diagonal terms $a_n$'s are also taken into consideration. 
\begin{align}
    \mathcal{L} |\mathcal{O}_n) =a_n|\mathcal{O}_n)+ b_n |\mathcal{O}_{n-1}) + b_{n+1} |\mathcal{O}_{n+1})\,. \label{tri2}
\end{align}
The differential equation corresponding to the expansion coefficients becomes\footnote{In the equation, diagonal elements $a_n$ are general. In our case, as they become purely imaginary, the last term in the RHS becomes $-\tilde{a}_n \phi_n(t)$.}
\begin{align}\label{diffeq2}
    \dot{\phi}_n(t) &= b_n \phi_{n-1}(t)-b_{n+1}\phi_{n+1}(t)+i a_n \phi_n (t) \\ 
    &= b_n \phi_{n-1}(t)-b_{n+1}\phi_{n+1}(t)- \tilde{a}_n \phi_n (t)\,,
\end{align}
where in the last line we replaced $a_n = i \tilde{a}_n$, where $\tilde{a}_n$ is real. The results and the orthonormality (of the order of $10^{-2}$) slightly improve for this case. We find that there exist purely imaginary diagonal coefficients in the tridiagonal representation of the Lindbladian matrix. The magnitude of the diagonal coefficients grows early and then reaches saturation. The behavior of the primary off-diagonal column $b_n$ also improves more than the fully Hermitian Lanczos algorithm. The $b_n$'s seem to differentiate between integrable and chaotic regimes initially. However, as one increases the bulk dephasing coupling $\gamma$, the growth of $b_n$ tends to deviate much earlier from their Hermitian counterparts. More on this and the plots are presented in the Appendix \ref{appa}. 

\begin{figure}[t]
   \centering
\begin{subfigure}[b]{0.46\textwidth}
\centering
\includegraphics[width=\textwidth]{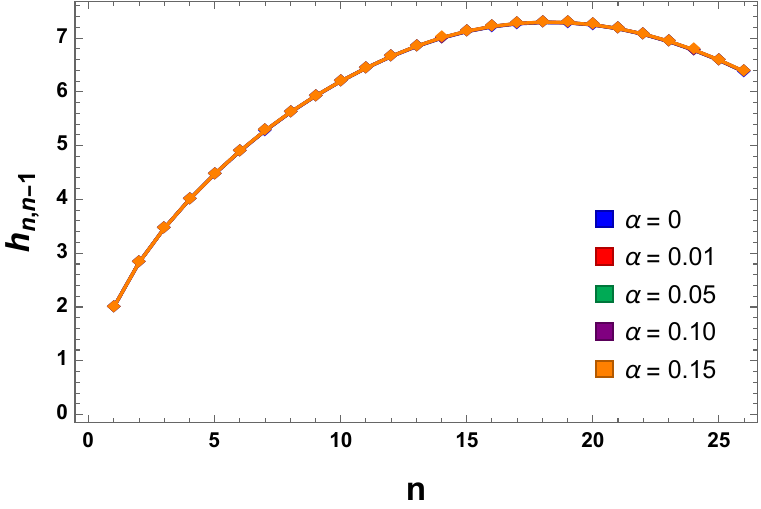}
\caption{}
\end{subfigure}
\hfill
\begin{subfigure}[b]{0.46\textwidth}
\centering
\includegraphics[width=\textwidth]{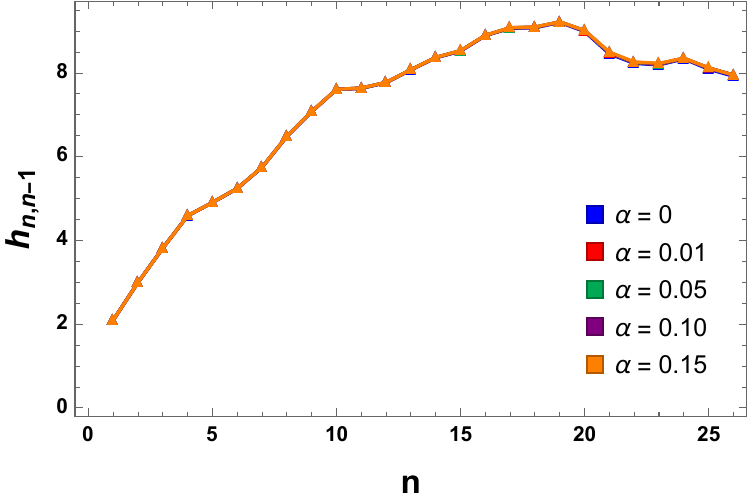}
\caption{}
\end{subfigure}
\hfill
\begin{subfigure}[b]{0.46\textwidth}
\centering
\includegraphics[width=\textwidth]{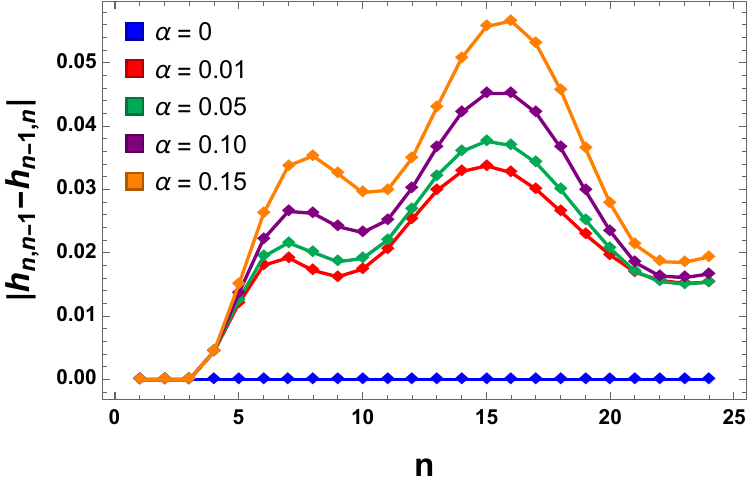}
\caption{}
\end{subfigure}
\hfill
\begin{subfigure}[b]{0.46\textwidth}
\centering
\includegraphics[width=\textwidth]{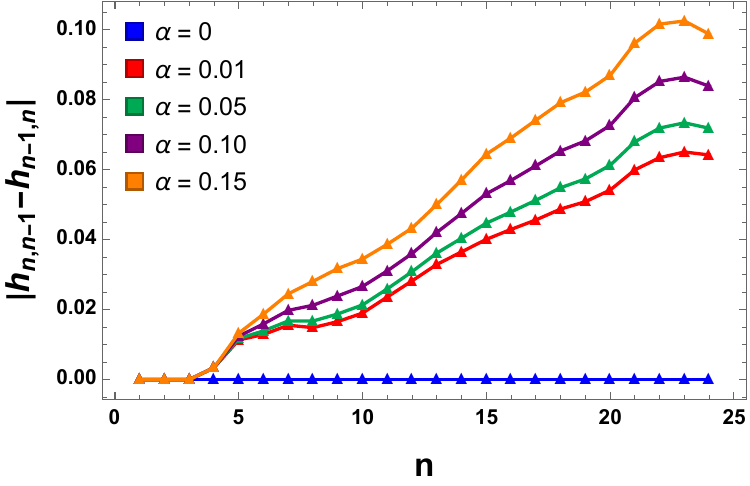}
\caption{}
\end{subfigure}
\caption{Growth of Arnoldi coefficients $h_{n,n-1}$ of the operator $Z_3$ in (a) integrable and (b) chaotic limits for bulk dephasing $\gamma = 0.1$ using Arnoldi iteration. Growth of the difference $|h_{n,n-1} - h_{n-1,n}|$'s of the operator $Z_3$ in (c) integrable and (d) chaotic limits for bulk dephasing $\gamma = 0.1$ using Arnoldi iteration. The blue curves are for closed system i.e., for $\alpha = 0$ and $\gamma = 0$. We choose $N=6$.} \label{fig:Arr2}
\end{figure}

\section{Arnoldi iteration}\label{arnoldi}
The obvious reason for this breakdown and inability to produce physically meaningful results for the Lanczos algorithm seems to be the presence of the non-Hermitian Lindbladian. It is not completely surprising as the original efficiency of the Lanczos algorithm stems from its ability to approximate the eigenvalues of a Hermitian matrix. This calls for an efficient method that extends beyond the Hermitian matrices.
A known method in literature is the Arnoldi iteration \cite{Arnoldi1951ThePO, Arnoldi2, Minganti:2021gxs} for approximating the eigenvalues of a non-Hermitian matrix. In fact, the Lanczos algorithm can be derived as a special case of the Arnoldi iteration when the matrix is symmetric and Hermitian. Instead of subtracting one previous vector as done in the Lanczos algorithm, in Arnoldi iteration, one performs the reduction of all previous vectors. 
More generally, a two-dimensional array is produced for the matrix form of the Lindbladian in the non-Hermitian case. Recall that the closed system Liouvillian ($\mathcal{L}_c \, \bullet=[H,\bullet]$) in Heisenberg picture has a tridiagonal form after implementation of the Lanczos algorithm with only $b_n$'s. The form is the following
\begin{equation}
    \mathcal{L}_{mn}^{(c)}  =(O_m|\mathcal{L}_c|O_n) ~~~\Rightarrow ~~~\mathcal{L}^{(c)} =\begin{pmatrix}
0 & b_1 & 0 & \cdots & 0\\
b_1 & 0 & b_2 & \cdots & 0\\
0 & b_2 & 0 & b_3 & \cdots\\
\cdots & \cdots & b_3 &\cdots &\cdots\\
0 & \cdots & \cdots &\cdots & b_n\\
0 & 0 & \cdots & b_n& 0\\
\end{pmatrix}\,.
\end{equation}
Hence, all the information is contained in the coefficients $b_n\in \mathbb{R}$.

\begin{figure}[t]
   \centering
\begin{subfigure}[b]{0.44\textwidth}
\centering
\includegraphics[width=\textwidth]{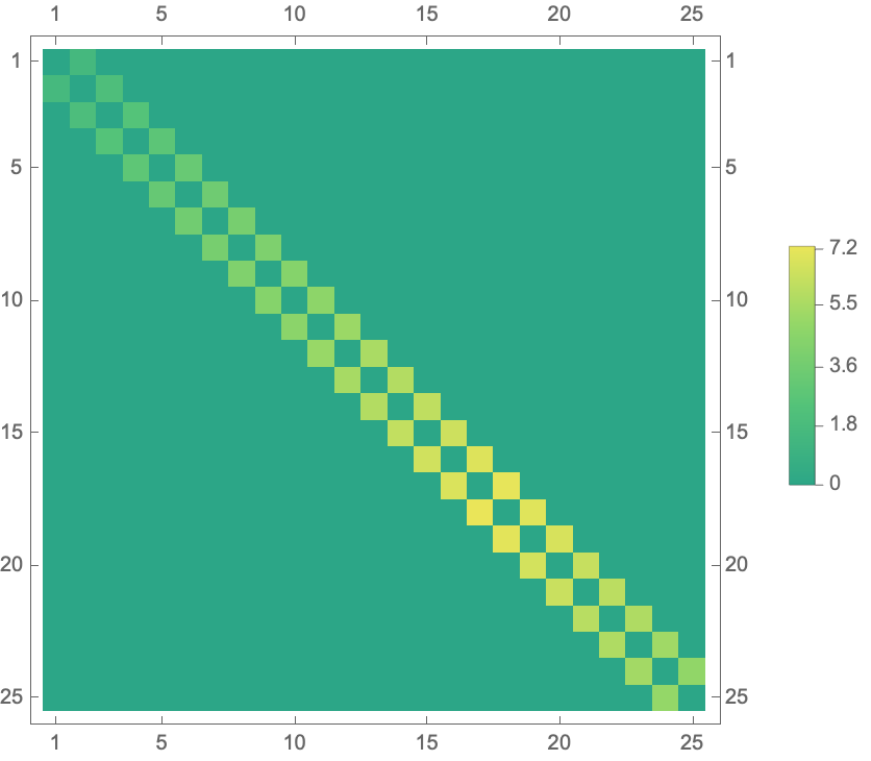}
\caption{}
\end{subfigure}
\hfill
\begin{subfigure}[b]{0.44\textwidth}
\centering
\includegraphics[width=\textwidth]{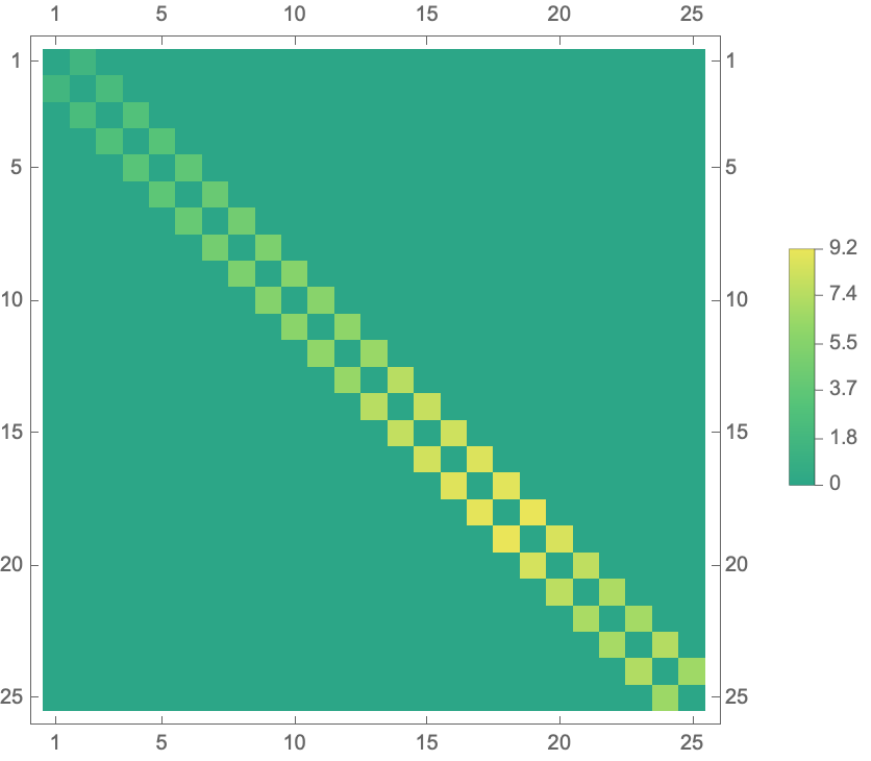}
\caption{}
\end{subfigure}
\hfill
\begin{subfigure}[b]{0.44\textwidth}
\centering
\includegraphics[width=\textwidth]{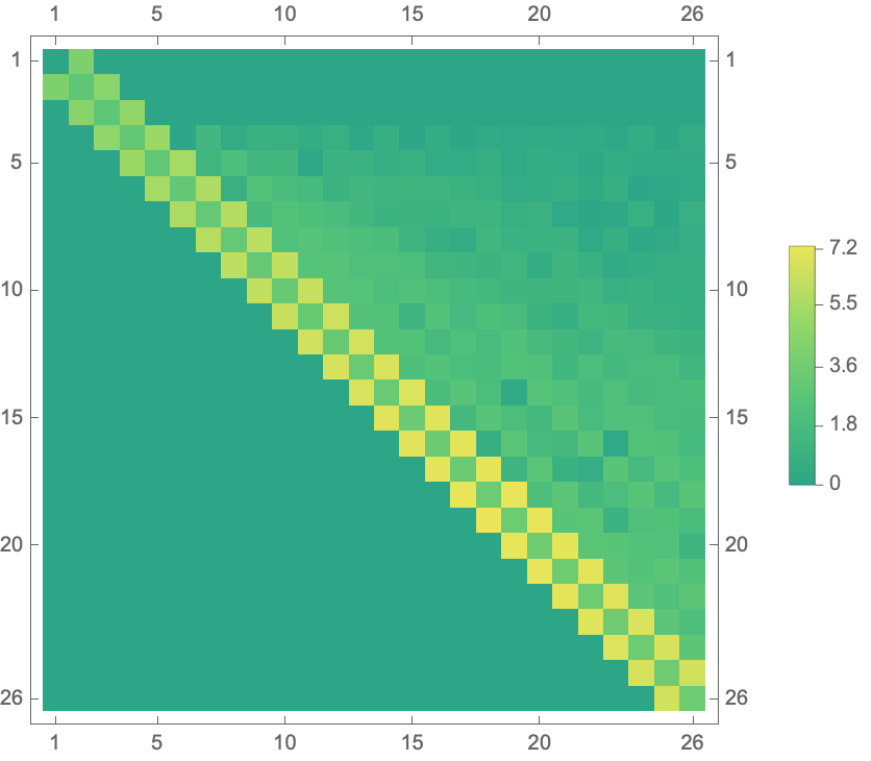}
\caption{}
\end{subfigure}
\hfill
\begin{subfigure}[b]{0.44\textwidth}
\centering
\includegraphics[width=\textwidth]{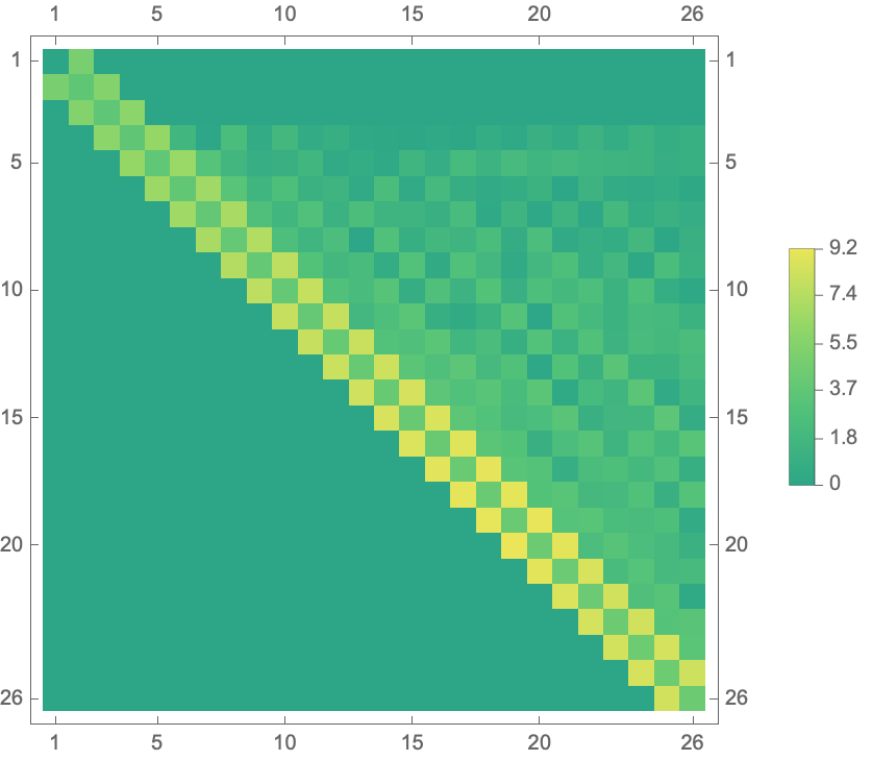}
\caption{}
\end{subfigure}
\caption{The Liouvillian ($\alpha = \gamma = 0)$ in tridiagonal form in (a) integrable and (b) chaotic limit. The Lindbladian in upper-Hessenberg form in (c) integrable and (d) chaotic limit for boundary Lindblad operators with $\alpha = 0.1$. We choose the lattice size $N=6$.} \label{fig:mat}
\end{figure}

On the other hand, the result of the Lindbladian for open systems by the implementation of the Arnoldi algorithm provides the following matrix form
\begin{equation}\label{Arnoldimatrix}
    \mathcal{L}_{mn}^{(o)}=\braket{v_m|\mathcal{L}_o|v_n} ~~~\Rightarrow ~~~\mathcal{L}^{(o)} =\begin{pmatrix}
h_{0,0} & h_{0,1} & h_{0,2} & \cdots & h_{0,n}\\
h_{1,0} & h_{1,1} & h_{1,2} & \cdots & h_{1,n}\\
0 & h_{2,1} & h_{2,2} & h_{2,3} & \cdots\\
\cdots & \cdots & h_{3,2} &\cdots &\cdots\\
0 & \cdots & \cdots &\cdots & h_{n-1,n}\\
0 & 0 & \cdots & h_{n,n-1} & h_{n,n}\\
\end{pmatrix}\,,
\end{equation}
where, the $\ket{v_n}$'s are the Arnoldi basis (see later) with $n = 0, 1, 2, \cdots$, Here the diagonal elements are purely imaginary while the primary off-diagonal elements are real. This matrix is in upper Hessenberg form i.e., $h_{i,j} = 0$ for $j \leq i-2$. The diagonal elements, again being purely imaginary, contribute to a decaying term as in \eqref{diffeq2}. This reflects the crucial fact that that the individual wavefunctions $\phi_n(t)$ decay with time as in $\sum_n |\phi_n(t)|^2$. This is a direct result of the system interacting with an environment.

The Arnoldi iteration for generating the two-dimensional array is the following. Let us consider a high dimensional sparse non-Hermitian square matrix (operator in this case) $\mathcal{L}_o$ and fix an initial normalized vector $|v_0\rangle$ on which the matrix acts as an operator. Then, the eigenvalues of $\mathcal{L}_o$ can be approximated by forming an iterative basis starting from $|v_0\rangle$ in the following steps, which has to be repeated for $k=1,2, \cdots,n$, where $n$ will be specified later.
\begin{align}
    |u_k\rangle= \mathcal{L}_o|v_{k-1}\rangle\,.
\end{align}
For $j=0$ to $k-1$, perform the following steps\footnote{The norm $||u_k||$ is defined as $||u_k|| = \sqrt{\braket{u_k|u_k}}$.}
\begin{align}
    &1. ~h_{j,k-1} = \langle v_j|u_k\rangle\,.~~~~~~~~~~~~~~~~~~~~~~~~ ~~~~~~~~~~~~ ~~~~~~~~~~~~~~~~~~~~~~~~~   \nonumber \\
    &2. ~|u_{k}\rangle  =|u_{k}\rangle-\sum_{j=0}^{k-1} h_{j,k-1} |v_{j}\rangle\,. \nonumber \\
    &3. ~ h_{k,k-1} =||u_k||\,.
\end{align}
Stop if $h_{k,k-1}=0$. Otherwise define $|v_k\rangle$ as
\begin{equation}
    |v_k\rangle= \frac{|u_k\rangle}{h_{k,k-1}}\,.
\end{equation}
It is easy to see that $h_{k,k-1}$, the norms of $|u_k\rangle$ are the analogues of $b_n$ in this iteration process.


Thus, we have constructed the entries of the array 
\begin{equation}
    h_{m,n}= \braket{v_m| \mathcal{L}_o|v_{n}}\,,
\end{equation}
with $m, n = 0, 1, 2, \cdots$. We call the coefficients $h_{m,n}$ as Arnoldi coefficients, as per the Arnoldi iteration. This constructs the upper triangular part of the Hessenberg matrix from the non-Hermitian matrix. The off-diagonal set of entries just below the diagonals, which can be termed as $h_{n,n-1}$, which are the norms of the vectors $u_n$.
In the newly formed basis, the Lindbladian superoperator is written as a $p \times p$ upper-Hessenberg matrix $\mathcal{H}_p$ such that \cite{Saad, Arn}
\begin{align}
    \mathcal{L}_o\, \mathcal{Q}_p = \mathcal{Q}_p \mathcal{H}_p + h_{p,p-1} \, v_p\, \xi_p^T\,, \label{uhess}
\end{align}
where $\mathcal{Q}_p = \{\ket{v_0}, \ket{v_1}, \cdots, \ket{v_{p-1}}\}$ is the set of Arnoldi basis vectors and $\xi_p^T = [0, 0, \cdots,1]^T$ is the $p$-th unit vector. For the Hermitian case, $\mathcal{H}_p$ reduces to the tridiagonal form of the Liouvillian with Krylov basis and \eqref{uhess} reduces to \eqref{tri}. The algorithm stops at $b_{\mathcal{K}}$, where $\mathcal{K}$ denotes the dimension of Krylov subspace. In such case, the Krylov basis provides an orthogonal restriction of the Lindbladian onto the $p$-th Krylov subspace, which can be seen as $\mathcal{Q}_p^{*} \mathcal{L}_o\, \mathcal{Q}_p = \mathcal{H}_p$. The eigenvalues of the upper Hessenberg matrix $\mathcal{H}_p$ equals the eigenvalues of the Lindbladian and they are known as Ritz values, corresponding to the eigenvectors known as Ritz vectors.

\begin{figure}[t]
   \centering
\begin{subfigure}[b]{0.46\textwidth}
\centering
\includegraphics[width=\textwidth]{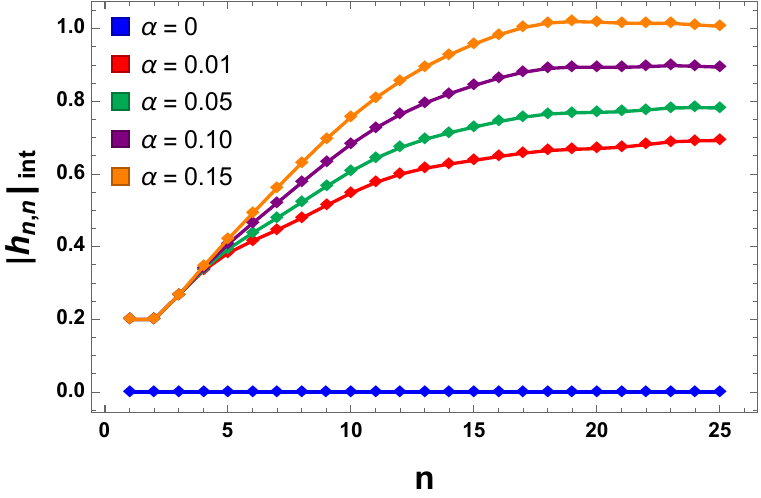}
\caption{}
\end{subfigure}
\hfill
\begin{subfigure}[b]{0.46\textwidth}
\centering
\includegraphics[width=\textwidth]{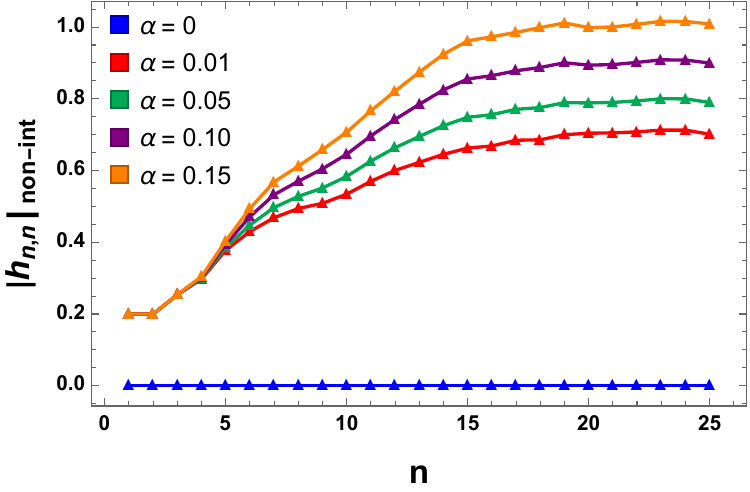}
\caption{}
\end{subfigure}
\hfill
\begin{subfigure}[b]{0.46\textwidth}
\centering
\includegraphics[width=\textwidth]{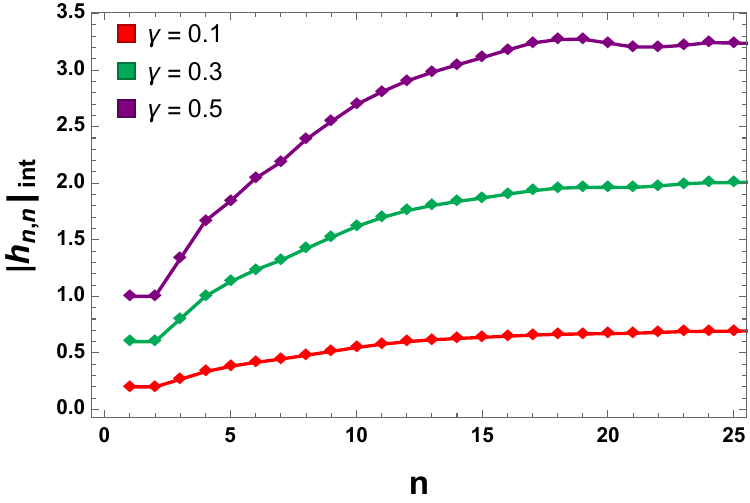}
\caption{}
\end{subfigure}
\hfill
\begin{subfigure}[b]{0.46\textwidth}
\centering
\includegraphics[width=\textwidth]{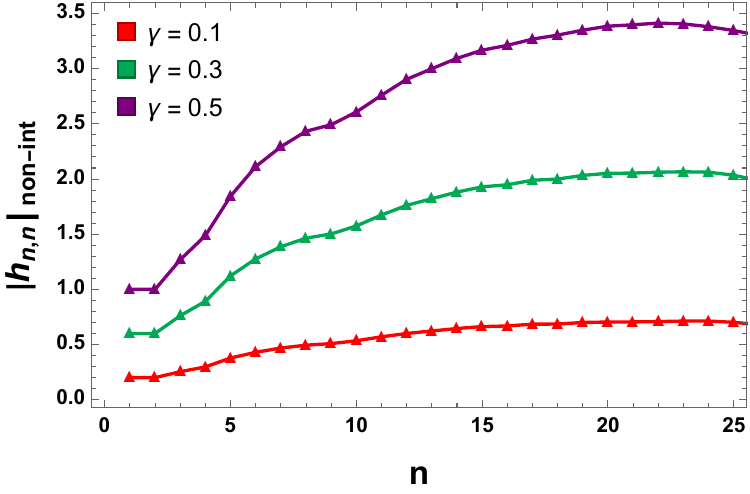}
\caption{}
\end{subfigure}
\caption{Growth of $|h_{n,n}|$'s of the operator $Z_3$ in (a) integrable and (b) chaotic limits for bulk dephasing $\gamma = 0.1$ using Arnoldi iteration. ($\gamma=0$, for $\alpha=0$) (c) integrable and (d) chaotic for boundary Lindblad $\alpha = 0.01$. We choose $N=6$. The blue curves are for closed system i.e., for $\alpha = 0$ and $\gamma = 0$.} \label{fig:Ardiagonals}
\end{figure}

Applying the Arnoldi iteration, we see that the situation drastically changes. The qualitative nature of the plots in Fig.\,\ref{fig:Arr2} (a) and (b) for $h_{n,n-1}$ with $n$ remains intact in the integrable and the non-integrable limit (see Appendix \ref{appb} for high bulk dephasing). Especially, we do not observe any unbounded growth with increasing dissipation, as we have observed implementing the Lanczos algorithm. Hence, to capture the true nature of a general system (whether integrable or non-integrable) via operator growth, we should perform the Arnoldi iteration instead of the Lanczos algorithm, where the latter fails to apprehend the nature of the system with increasing non-Hermiticity. However, the $h_{n,n-1}$'s in the case of Arnoldi iteration fail to show the change of non-Hermiticity parameters $\alpha$ and $\gamma$. Effectively, the behavior of the norms remains the same for all different non-Hermiticity parameters.

\begin{figure}[t]
   \centering
\begin{subfigure}[b]{0.46\textwidth}
\centering
\includegraphics[width=\textwidth]{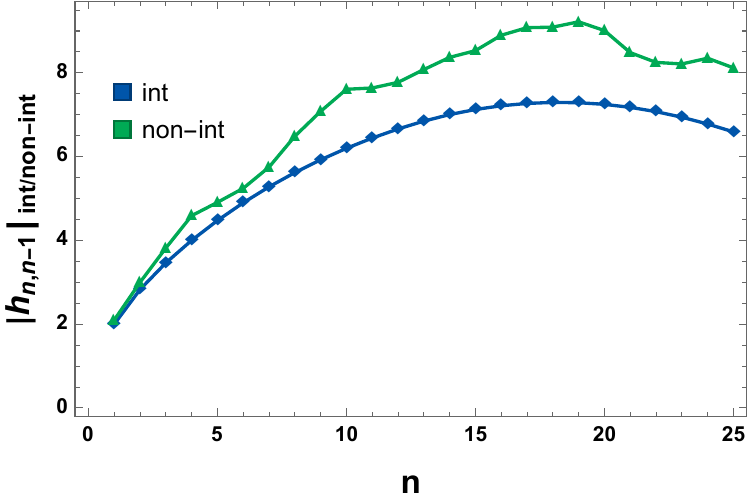}
\caption{}
\end{subfigure}
\hfill
\begin{subfigure}[b]{0.46\textwidth}
\centering
\includegraphics[width=\textwidth]{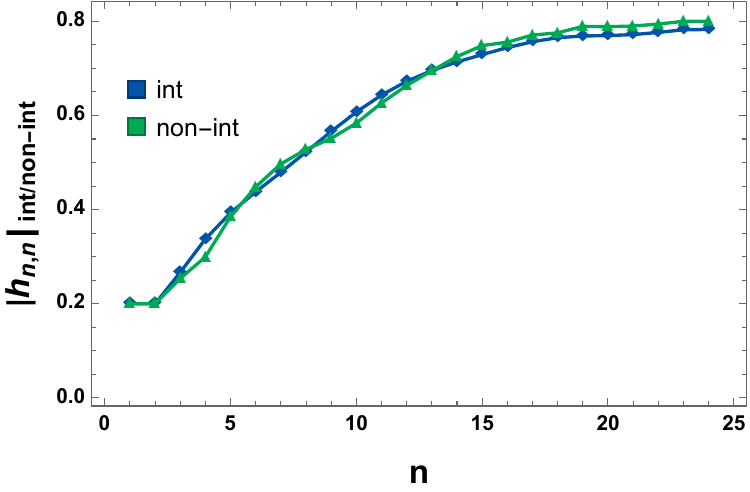}
\caption{}
\end{subfigure}
\hfill
\begin{subfigure}[b]{0.46\textwidth}
\centering
\includegraphics[width=\textwidth]{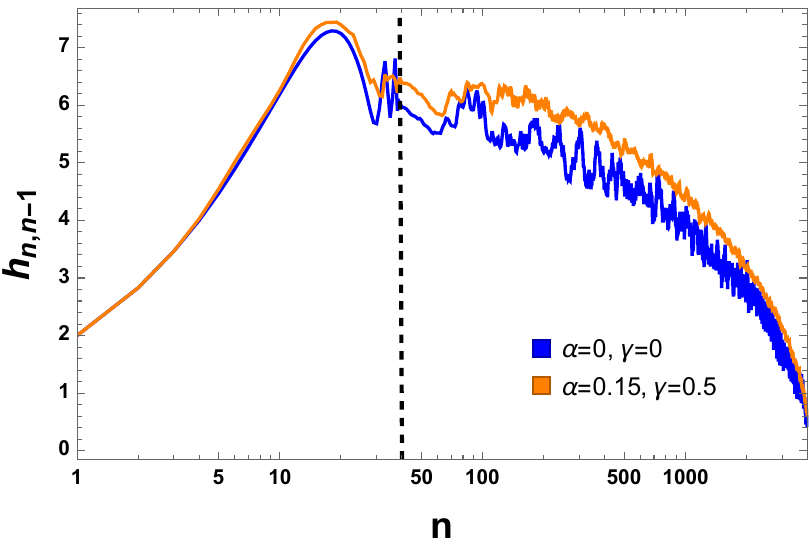}
\caption{}
\end{subfigure}
\hfill
\begin{subfigure}[b]{0.46\textwidth}
\centering
\includegraphics[width=\textwidth]{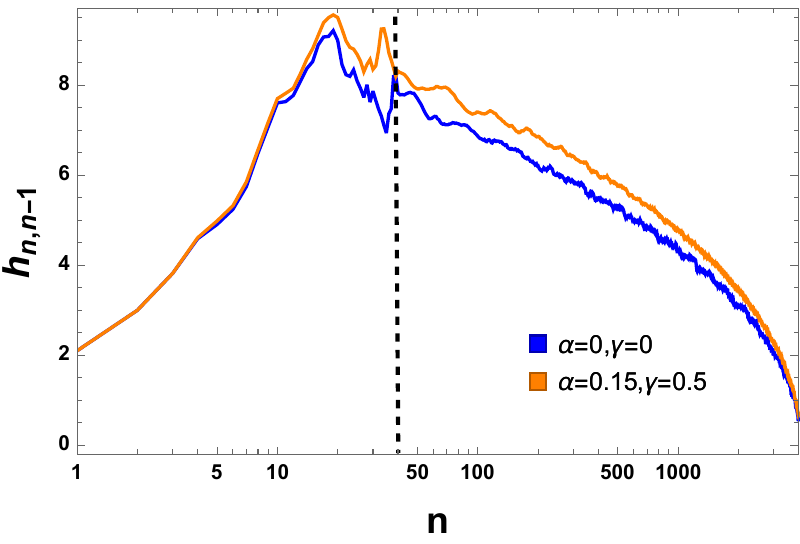}
\caption{}
\end{subfigure}
\caption{Comparison between integrable and chaotic regimes from (a) growth of $h_{n,n-1}$'s (distinguishable)  and (b) growth of $|h_{n,n}|$'s (indistinguishable) of the operator $Z_3$ by implementing the Arnoldi algorithm with $\gamma=0.1$ and $\alpha=0.05$. $h_{n,n-1}$ vs $n$ logarithmic plot of the operator $Z_3$ for $n$ up to the Krylov space dimension (for our example, it is around $\mathcal{K} = (D^2-D+1) \approx 4 \times 10^{3}$ (with $D = 2^6$) for both integrable and chaotic regimes) with Arnoldi iteration in (c) integrable and (d) chaotic limits. The fluctuation averaging is done after $n>40$ (indicated by the dashed line). We choose $N=6$. The blue curves are for closed system i.e., for $\alpha = 0$ and $\gamma = 0$.} \label{fig:Ardiagonalsi}
\end{figure}

However, there are supposed to be other entries in the Hessenberg form than just the primary off-diagonal entries\footnote{We thank Xiangyu Cao for bringing our attention to this point which clarifies a confusion in the preliminary version of the manuscript.}. They are much smaller compared to the tridiagonal elements. However, they are not so small that they can be ignored. To be more precise, the coefficients $h_{n,n+2m-1}$ for integer $n$ and $m$ are all real. On the other hand, coefficients $h_{n,n+2m}$ are all purely imaginary. Hence if one looks at the consecutive diagonal series, they are alternating real and purely imaginary. However, the coefficients apart from the ones focussed in this article are of the order of $10^{-1}$. Finally, the orthogonality for the Arnoldi case is far better than the Lanczos and is of the order of $10^{-10}$ or even less. Hence, the vectors constructed in this procedure can be treated as forming a completely orthonormal complete basis. However, solving for the wavefunctions $\phi_n(t)$'s exactly becomes harder in this case as there are many more coefficients involved in the differential equation.

We numerically observe that $h_{n,n-1} \neq h_{n-1,n}$. In fact, their differences behave non-linearly with $n$ for both in integrable and chaotic limit (see Fig.\,\ref{fig:Arr2} (c), (d)). Moreover, the other non-diagonal elements also do not vanish (except the first three rows) although their magnitude is far less compared to the primary of off-diagonal elements. In
Fig.\,\ref{fig:mat}, we have shown the matrix elements in tridiagonal (for closed system) and the upper-Hessenberg form. The elements vary significantly with the increase of bulk dephasing parameter (see Appendix \ref{appb}).
These two features are completely new for the Lindbladian and in sharp contrast with the usual Liouvillian which takes a tridiagonal form in Krylov basis.
The purely imaginary diagonal entries $h_{n,n}$ are a clear reflection of non-Hermiticity in the case of open systems. When the magnitude of the diagonal entries is plotted with $n$ (see Fig.\,\ref{fig:Ardiagonals}), they successfully capture the information about different non-Hermiticity parameters $\alpha$ and $\gamma$.\footnote{Note that the diagonal entries are zero for closed system, $\alpha=0,\, \gamma=0$. Hence, both the Lanczos algorithm and Arnoldi iteration give the same results in this case.} However, when the growth of the magnitude of the diagonal entries for integrable and chaotic cases are compared, they are indistinguishable (see Fig.\,\ref{fig:Ardiagonalsi} (a), (b)).

\textbf{More on the form of the Lindbladian matrix:} Here we would like to focus on the form of the Arnoldi matrix. Firstly,  $h_{n,n-1}\neq h_{n-1,n}$ is just an artefact of the construction of the Arnoldi matrix. In the usual generalized Lanczos algorithm, since we anyway consider only two sets of coefficients $a_n$ and $b_n=h_{n-1,n}$, we ignore any possible differences between the row below the diagonals ($h_{n,n-1}$). In the Arnoldi matrix as well, if we look more closely, this difference comes only when there are nonzero coefficients in the upper triangular matrix other than the tridiagonal part (for example, $3$rd or $4$th off-diagonal rows and columns in Fig.\,\ref{fig:mat} (c) and (d)). Hence the difference can be attributed to the presence of other matrix elements in the Arnoldi form. Effectively, this difference spreads in the other set of coefficients. It is worth mentioning that if one makes a systematic approximation of obtaining these small extra coefficients, then an effective tridiagonal form will reappear. For the moment, we ignore them (set them to zero) and we will get back to them in Appendix \ref{appc}.
\begin{enumerate}
    \item \textbf{Real primary off-diagonal elements $(h_{n,n-1}$ and $h_{n-1,n})$:} The coefficients $h_{n,n-1}$ and $h_{n-1,n}$ are the norms of vectors $u_n$ as explained in the algorithm. These norms will always be real numbers since even for a non-Hermitian vector $u_n$; the norm is defined as $\mathrm{Tr}(u_n^{\dagger}u_n)$, which can only be a real positive number. 
    
    \item \textbf{Purely imaginary diagonal elements $(h_{n,n})$:} Now that we know the $h_{n,n-1}$ or $b_n$-like elements are real, we might expect some complex numbers to show up in the diagonals. However, as we have seen in earlier sections these elements are purely imaginary. Then it is a valid question to ask why do not these coefficient have a real part. To address this issue, we note that in previous literature \cite{Albert_2014}, the Lindbladian evolution has been studied in the form $e^{\tilde{\mathcal{L}_o}t}$. In these cases, the Lindbladian eigenvalues are found to always appear in real and complex conjugate pairs. The complex conjugate pair of eigenvalues result from the non-unitary evolution and are extremely crucial in characterizing the non-unitary evolution. Now for this to hold, the polynomial characteristics equation of the Lindbladian matrix $|\tilde{\mathcal{L}}_o-\lambda I|=0$ form must satisfy the complex conjugate root theorem \cite{croot}, which says that only a polynomial equation with all real coefficients can have either real or complex conjugate pair as roots. Now, since we considered the evolution as $e^{i\mathcal{L}_o t}$, we simply have to multiply our tridiagonal matrix form (assuming the other terms are just the difference between $h_{n,n-1}$ and $h_{n-1,n}$) to $i$ and then look at the properties of the polynomial characteristics equation. This modified matrix will therefore have purely negative and real diagonals and purely imaginary $h_{n,n-1}$. The form of the matrix will be
    \begin{equation}\label{modifiedL}
        \tilde{\mathcal{L}_o}=i \mathcal{L}_{o} =\begin{pmatrix}
-|h_{1,1}| & i h_{1,2} & 0 & \cdots & 0\\
i h_{2,1} & -|h_{2,2}| & i h_{2,3} & \cdots & 0\\
0 & i h_{3,2} & -|h_{3,3}| & i h_{3,4} & \cdots\\
\cdots & \cdots & i h_{4,3} &\cdots &\cdots\\
0 & \cdots & \cdots &\cdots & i h_{n-1,n}\\
0 & 0 & \cdots & i h_{n,n-1}& -|h_{n,n}|\\
\end{pmatrix}\,.
    \end{equation}
For this matrix, it can be shown that all the coefficients of the characteristics polynomial equation are real. On the other hand, the possibility of a real part in the diagonals of our $\mathcal{L}_o$ will result in an imaginary part in the diagonals of the modified $\tilde{\mathcal{L}_o}$ (of the form of $i h_{n,n}= -\mathrm{Im}[h_{n,n}]+ i \mathrm{Re}[h_{n,n}]$). In that case, we find that inevitably the characteristics equation will have complex coefficients.\footnote{We thank Prashanth Raman for helping us in the sketch of this argument.} Therefore, by the complex conjugate roots theorem, this can never give real and complex conjugate pairs of eigenvalues. We give more detailed arguments in the Appendix \ref{appc}. With these arguments, we can conclude that the upper Hessenberg form of the Lindbladian with alternating real and purely imaginary values is unique and necessary for producing proper eigenvalues. Hence, this particular nature of alternating diagonals will be independent of the choice of Hamiltonian and the choice of coupling, and hence they are generically true for dissipative quantum systems. 
\end{enumerate}

When the Krylov basis is explored completely, the Arnoldi coefficients hit zero. See Fig.\,\ref{fig:Ardiagonalsi} (c), (d). This has to be the case, as we have taken a finite-dimensional system. There are a couple of points to remember for the case of a fully explored Krylov basis. In \cite{Rabinovici:2020ryf}, the authors found that for the integrable SYK model, the Krylov dimension is much smaller than the chaotic case. However, as argued in \cite{Rabinovici:2021qqt}, this is not true for generic integrable cases where the Krylov space dimension appears to be the same for both integrable and chaotic cases. Due to the phenomenon dubbed the Krylov-space localization \cite{Rabinovici:2021qqt}, the space is explored less efficiently when the integrable yet strongly interacting XXZ spin chain is analyzed. This localization results in the saturation of Krylov complexity for integrable XXZ chain at a lower value than the chaotic case and happens in the fully symmetry-reduced sector.\footnote{We thank Julian Sonner and the anonymous referee for pointing out these facts to us.}  In our case, we find the Krylov space dimensions for integrable and chaotic cases to be almost the same, which is in agreement with the results of \cite{Rabinovici:2021qqt}. It is worth mentioning that we find many fluctuations for higher Arnoldi coefficients, which can be averaged for the $s$ number of preceding and following coefficients for a specific coefficient (see caption of Fig.\,\ref{fig:Ardiagonalsi} more details). After this fluctuation-averaging, the behavior of the Arnoldi coefficients becomes much smoother, and one can observe the full exploration of the Krylov basis much more clearly. Yet, we can see that for integrable systems, the coefficients fluctuate more rapidly than the chaotic ones, which, according to \cite{Rabinovici:2021qqt}, can again be attributed to the Krylov space localization for the integrable case. However, the non-Hermitian effects tend to reduce it in both integrable and chaotic limits. At this stage, we do not have a clear understanding of it, and it requires further investigation.

\section{Conclusion and outlook}\label{conc}
Dissipation and decoherence are ubiquitous in nature; hence it is of primary interest to study the operator growth in such systems. Our results indicate that the Lanczos algorithm, while well-suited for Hermitian operators, breaks down if the superoperator becomes non-Hermitian, e.g., in the case of open system Lindbladian evolution. The idea of information scrambling under dissipation also becomes problematic since the coefficients never saturate and exhibit unbounded growth. With changing the coupling with the environment, the growth rate of Lanczos coefficients changes (as discussed in Appendix \ref{appa}), and different kinds of interactions can be distinguished. On the other hand, the Arnoldi iteration, when applied to a non-Hermitian system, distinguishes both integrable and chaotic regimes as well as different non-Hermiticity parameters. The norms $h_{n,n-1}$, analogous to $b_n$, distinguish the integrable and chaotic regimes, whereas the purely imaginary diagonal entries $h_{n,n}$ capture the information about the non-Hermiticity parameters.\footnote{The diagonal entries also appear in the state complexity picture (although real in that case), where Liouvillian is replaced by the Hamiltonian \cite{Balasubramanian:2022tpr, Caputa:2022eye}.} The Liouvillian takes an upper-Hessenberg form which is not symmetric in its elements, a clear contrasting feature of tridiagonal Liouvillian of a closed system. This work, therefore, clearly reflects the breakdown of the Hermitian methods in open systems and supports the application of non-Hermitian methods like Arnoldi iteration.

There are a few interesting questions that need to be understood. Firstly, it would be interesting to back the conclusions by studying other quantum mechanical, and field-theoretic models like open XXZ chain \cite{PhysRevX.10.021019} or open Sachdev-Ye-Kitaev model \cite{Sa:2021tdr, Kulkarni:2021gtt} with corresponding Lindblad operators. One might test the validity of UOGH in open quantum field theory \cite{Baidya:2017eho}, i.e., whether the asymptotic scaling of the primary off-diagonal Arnoldi coefficients is linear in $n$, similar to the Lanczos coefficients for the closed systems \cite{Dymarsky:2021bjq}. The interesting question is, what is the asymptotic scaling of (imaginary) diagonal elements? 
From the numerical examples, we see that their saturation depends on the interaction with the environment. Do they have any description in terms of a Toda chain flow \cite{Dymarsky:2019elm}? Further, the upper-Hessenberg form of the Lindbladian asks for the role of the other off-diagonal elements than the primary ones. Especially, the smallness of their magnitude seeks whether they can be potentially eliminated by using repeated re-orthogonalization \cite{Rabinovici:2020ryf}. However, we do not have an analytic result at this stage and would like to return to this issue in the future.

Although here we have only restricted to the growth of Lanczos and Arnoldi coefficients which serves our purpose, it would be interesting to compute Krylov complexity in such cases where the comparison with circuit complexity on open systems \cite{Bhattacharyya:2021fii} could be possible. The primary obstacle is to ensure the conservation of probability on the Krylov basis, which might be lost, and show decay over time. This is potentially due to the presence of purely imaginary non-zero diagonal elements, even if they are small compared to the primary off-diagonal elements. One practical way could be to implement by bi-orthogonalizing the Lanczos algorithm \cite{Gruning, bilan}, which will render the Lindbladian in pure tridiagonal form with different sets of coefficients.\footnote{We thank Xiangyu Cao for discussions on this point.} Computing complexity, in such cases, could be practically possible without encountering significant numerical errors.


Our analysis opens up possibilities for the exploration of a vast field of non-Hermitian Hamiltonians, especially with having the PT-symmetry \cite{Bender:1998ke, Bender:2007nj}, which has gained significant importance in recent experiments \cite{Ruter2010, feng, Assawaworrarit2017}. By the balanced gain and loss mechanism, the closed PT symmetric system can be described by an open system dynamics such that the evolution of density matrix reads \cite{10.21468/SciPostPhys.9.4.052, PhysRevA.105.022219, broody, Cornelius:2021ccu}
\begin{align*}
    \dot{\rho} = - i \big(H_{\mathrm{eff}} \rho - \rho H_{\mathrm{eff}}^{\dagger} \big), ~~~~ H_{\mathrm{eff}} = H - \frac{i}{2} \sum_i L_i^{\dagger} L_i \,,
\end{align*}
where $H_{\mathrm{eff}}$ is the \emph{effective} non-Hermitian Hamiltonian made up from the Hermitian Hamiltonian $H$ and the Lindblad operators $L_i$. In other words, this suggests that the operator growth in such systems will be controlled by $H_{\mathrm{eff}}$, which can efficiently obtained by implementing the Arnoldi iteration. We wish to report such issues in future studies.

\section*{Acknowledgements}
We wish to thank Hugo A. Camargo, Xiangyu Cao, Pawel Caputa, Shajid Haque, Arnab Kundu, Sabyasachi Maulik, Jeff Murugan, Tatsuma Nishioka, Prashanth Raman, Sayak Ray, Shibaji Roy, Aninda Sinha, Masaki Tezuka, and Hendrik J.R. van Zyl for illuminating discussions and helpful comments on the draft. PN thanks Xiangyu Cao for ongoing collaboration on related topics and Tanay Pathak for implementing the numerical methods in \emph{Mathematica}. PN would like to thank NITheCS and the University of Cape Town for the warm hospitality during the final stages of the work, where parts of the results were presented. AB would like to thank the TP III division of the University of W\"{u}rzburg, and the faculty of Physics, University of Warsaw, for hosting him during the final stages of the work. AB is supported by the Institute of Eminence endowed postdoctoral fellowship offered by the Indian Institute of Science. The work of PN is supported by the JSPS Grant-in-Aid for Transformative Research Areas (A) ``Extreme Universe'' No. 21H05190.

\appendix
\section{Appendix: More on operator growth and dissipation} \label{appa}

\begin{figure}[t]
   \centering
\begin{subfigure}[b]{0.46\textwidth}
\centering
\includegraphics[width=\textwidth]{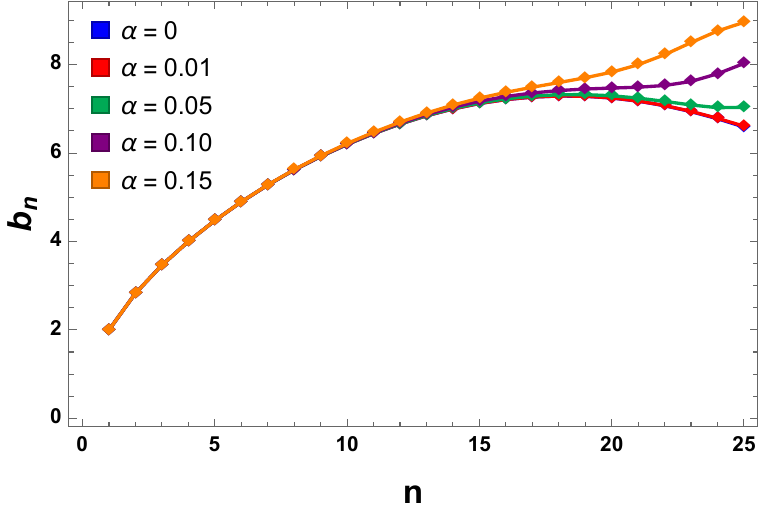}
\caption{}
\end{subfigure}
\hfill
\begin{subfigure}[b]{0.46\textwidth}
\centering
\includegraphics[width=\textwidth]{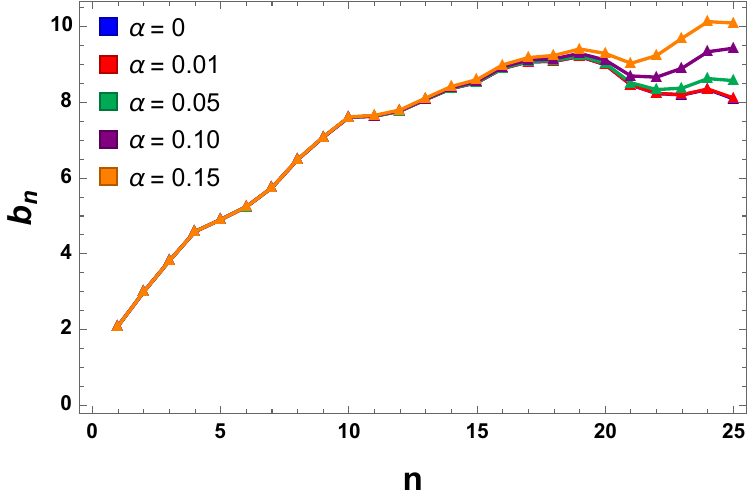}
\caption{}
\end{subfigure}
\hfill
\begin{subfigure}[b]{0.46\textwidth}
\centering
\includegraphics[width=\textwidth]{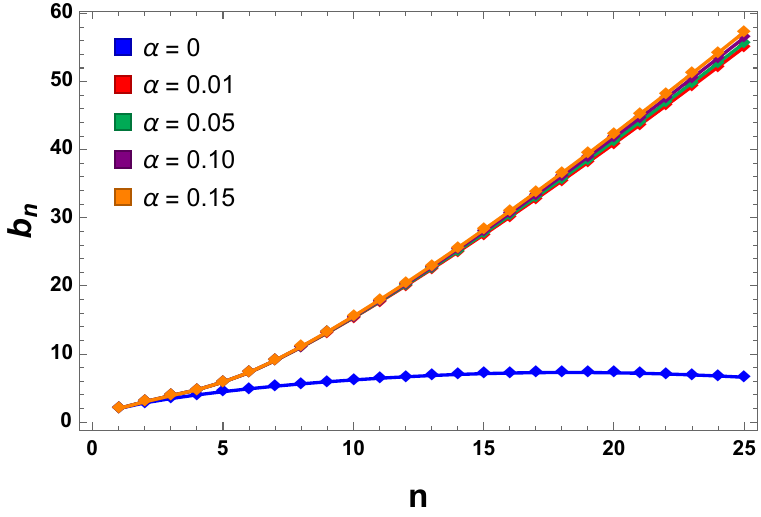}
\caption{}
\end{subfigure}
\hfill
\begin{subfigure}[b]{0.46\textwidth}
\centering
\includegraphics[width=\textwidth]{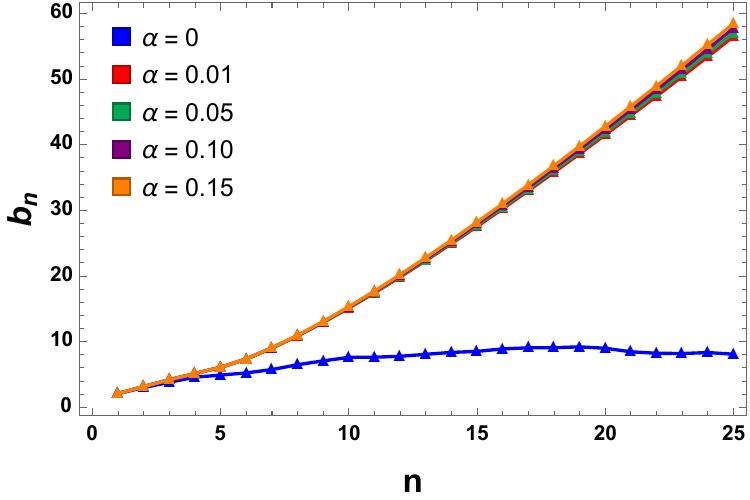}
\caption{}
\end{subfigure}
\caption{Growth of $b_n$'s of the operator $Z_3$ in (a) integrable and (b) chaotic limits by implementing the Lanczos algorithm with no bulk dephasing. Growth of $b_n$'s of the operator $Z_3$ in (c) integrable and (d) chaotic limits by implementing the Lanczos algorithm with high ($\gamma = 0.5$) bulk dephasing. We choose $N=6$. The blue curves are for closed system i.e., for $\alpha = 0$ and $\gamma = 0$.} \label{fig:Lan}
\end{figure}

In this Appendix, we show the different growth rates for different non-Hermiticity parameters. As explained in the main text, we have considered two kinds of Lindblad jump operators, the boundary amplitude damping ones with the corresponding coupling $\alpha$ and the boundary as well as bulk dephasing operators with the associated coupling $\gamma$. For the Lanczos algorithm, we show results for the following two cases, i) without considering diagonal coefficients and ii) including the diagonal coefficients.
\subsection{Lanczos without diagonals}
For Lanczos without considering diagonal coefficients, we show the plots for high bulk dephasing and without bulk dephasing. We compare the integrable and chaotic regimes of coupling parameters for different $\alpha$. This means we are varying the interaction strength between the open system and its environment. With more interaction, the non-Hermiticity parameters become more robust and affect the Lanczos iteration much earlier than when the coupling is weaker. The deviation from the Hermiticity (the commutator term in the Lindbladian) becomes more evident. In the main text, we provided the Lanczos plots for $\gamma=0.1$ with varying $\alpha=0.01,\, 0.05,\, 0.10,\,\text{and}\, 0.15$. Of course, in all of the comparison plots, we also have the closed system case as $\alpha=\gamma=0$.

In Fig.\,\ref{fig:Lan}, we provide the plots with varying coupling strengths of the dephasing interaction. We observe from Fig.\,\ref{fig:Lan} (a) and Fig.\,\ref{fig:Lan} (b) that, without bulk dephasing ($\gamma=0$) for all $\alpha$'s, the deviation of Lanczos coefficients from the closed system takes place at a larger $n$ value. This is because our operator is at the third site, and with no dephasing, the only Lindblad operators are placed in the boundary sites. Hence, they can only affect the evolution of the operator under study once the operator support reaches the boundary. However, for larger $\alpha$, the deviation from closed results occurs earlier than the smaller ones. At this point, therefore, we can clearly see the effect of how increasing non-Hermiticity in the boundary affects the operator growth.

\begin{figure}[t]
   \centering
\begin{subfigure}[b]{0.46\textwidth}
\centering
\includegraphics[width=\textwidth]{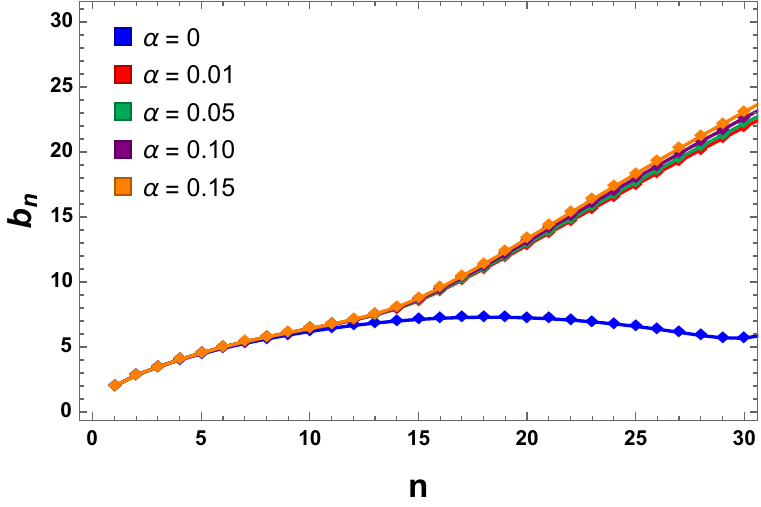}
\caption{}
\end{subfigure}
\hfill
\begin{subfigure}[b]{0.46\textwidth}
\centering
\includegraphics[width=\textwidth]{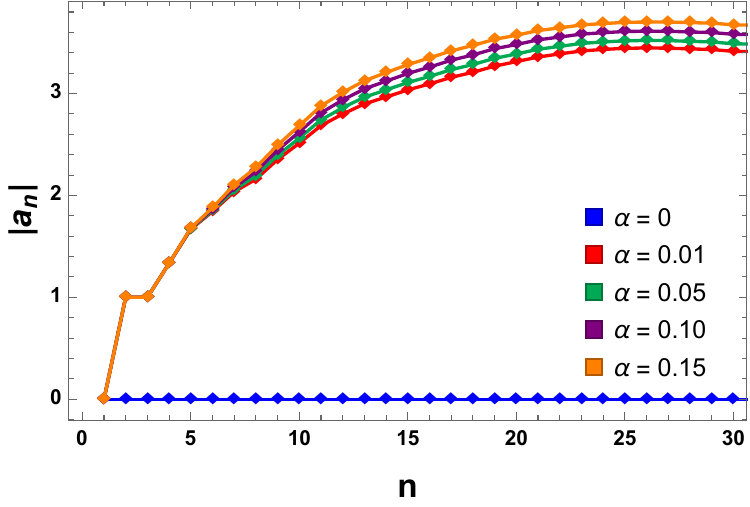}
\caption{}
\end{subfigure}
\hfill
\begin{subfigure}[b]{0.46\textwidth}
\centering
\includegraphics[width=\textwidth]{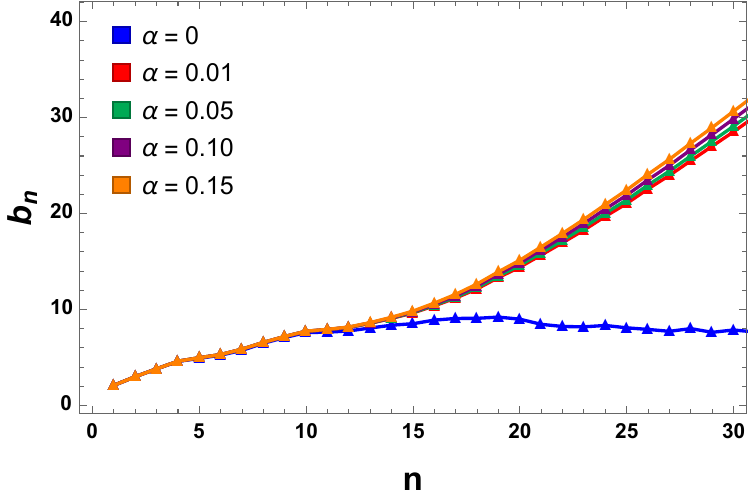}
\caption{}
\end{subfigure}
\hfill
\begin{subfigure}[b]{0.46\textwidth}
\centering
\includegraphics[width=\textwidth]{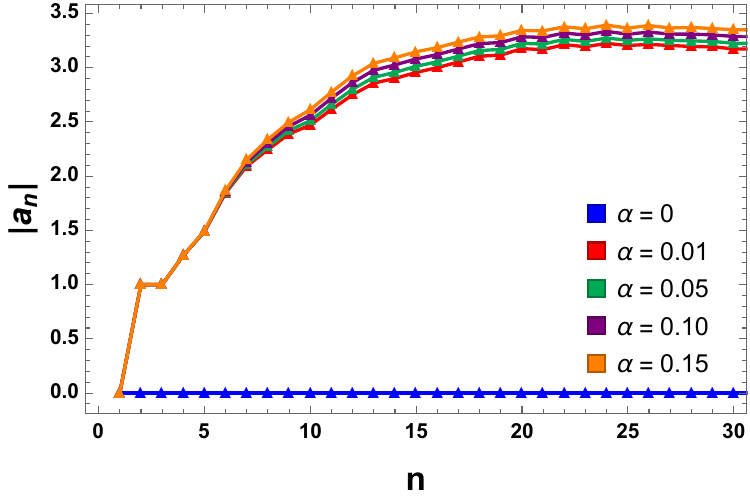}
\caption{}
\end{subfigure}
\caption{Growth of $b_n$'s ((a) and (c)) and $|a_n|$'s ((b) and (d)) of the operator $Z_3$ in integrable ((a) and (b)) and chaotic ((c) and (d)) limits by implementing the generalized Lanczos algorithm (with considering diagonal coefficients) with  high ($\gamma = 0.5$) bulk dephasing. We choose $N=6$. The blue curves are for closed system i.e., for $\alpha = 0$ and $\gamma = 0$.} \label{fig:gdLan}
\end{figure}

As shown in Fig.\,\ref{fig:Lan} (c) and (d), the deviation happens a lot earlier, much before the finite-size effects start kicking in, when bulk dephasing is stronger ($\gamma=0.5$). We observe that the coefficients for all the nonzero $\alpha$ cases deviate from $\alpha=0,\,\gamma=0$ (closed) results. This is simply because the strong dephasing effects dominate in this case, and since they are at all sites, including the one where the nontrivial part of the initial operator is situated. Even in this case, the different non-zero values of $\alpha$'s are clearly indistinguishable. 
These plots, therefore, strengthen our conclusions about the non-Hermiticity effects in Lanczos coefficients. Another interesting point to note here is that although Lanczos coefficients become insensitive to the integrability of the system, they tend to capture the information about different kinds of interaction strengths sensibly.

\subsection{Lanczos with diagonals}

In this case, we find that if we use the more general version of the Lanczos algorithm, where one can also consider diagonal components in the tridiagonal matrix form of the Lindbladian, the orthogonality improves slightly. While for the Hermitian Lanczos, typically used for a closed system, the orthonormality was of the order of $10^{-1}$, for the generalized Lanczos, it is found to improve to at least $10^{-2}$. The diagonal components are found to be purely imaginary. In this form of the algorithm, the matrix is of the following form
\begin{equation}
    \mathcal{L}_{mn}  =(O_m|\mathcal{L}|O_n) ~~~\Rightarrow ~~~\mathcal{L} =\begin{pmatrix}
i |a_0| & b_1 & 0 & \cdots & 0\\
b_1 & i |a_1| & b_2 & \cdots & 0\\
0 & b_2 & i |a_2| & b_3 & \cdots\\
\cdots & \cdots & b_3 &\cdots &\cdots\\
0 & \cdots & \cdots &\cdots & b_n\\
0 & 0 & \cdots & b_n& i |a_{n}|\\
\end{pmatrix}\,.
\end{equation}

\begin{figure}[t]
   \centering
\begin{subfigure}[b]{0.46\textwidth}
\centering
\includegraphics[width=\textwidth]{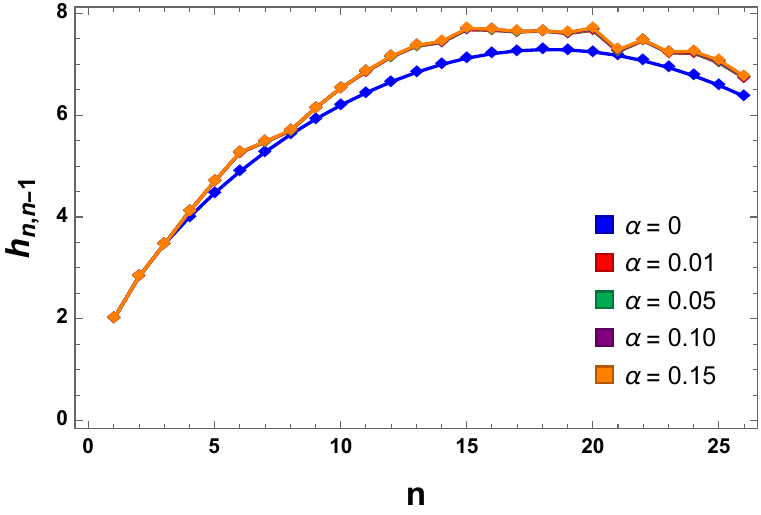}
\caption{}
\end{subfigure}
\hfill
\begin{subfigure}[b]{0.46\textwidth}
\centering
\includegraphics[width=\textwidth]{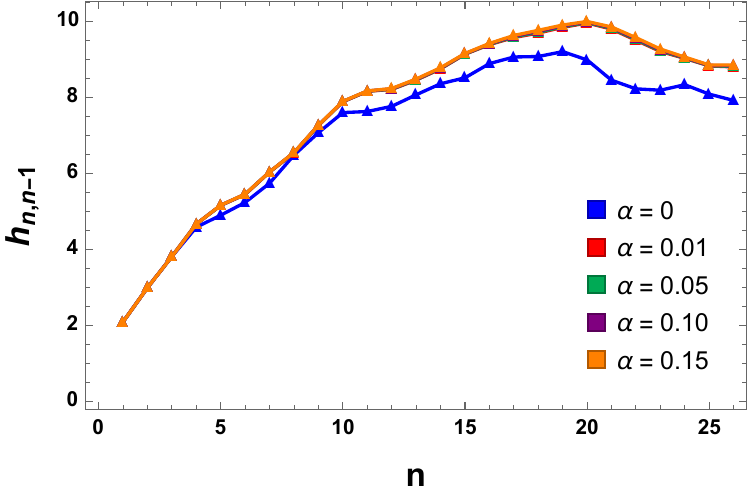}
\caption{}
\end{subfigure}
\caption{Growth of Arnoldi coefficients $h_{n,n-1}$'s of the operator $Z_3$ in (a) integrable and (b) chaotic limits for bulk dephasing $\gamma = 1$ using Arnoldi iteration. We choose lattice size $N=6$. Notice that for small bulk dephasing, the plots do not deviate from the closed system results (blue curve) corresponding to $\alpha = \gamma = 0$. For very large bulk dephasing, the plots deviate a little bit overall, but the plots can not still distinguish different $\alpha$s. The integrable and chaotic scaling remain distinguishable for all cases.} \label{fig:high}
\end{figure}
However, the orthonormality is still not satisfactory, especially if compared to the Arnoldi results. As shown in Fig.\,\ref{fig:gdLan} (a) and (c), although the $b_n$'s tend to distinguish between integrable and chaotic regimes better compared to the Hermitian Lanczos, they divert from the closed system results soon for high bulk dephasing. They also keep growing forever without ever showing a full exploration of the Krylov basis. This behavior of $b_n$ is problematic, as explained before. Another point to note here is that although the $b_n$'s keep growing forever, the growth rate for the chaotic case is more than the integrable. On the other hand, the Fig.\,\ref{fig:gdLan} (b) and (d) shows the growth of the diagonal coefficients $a_n$ (purely imaginary). The growth of $a_n$'s for both integrable and chaotic are similar. The distinguishability is, therefore, somewhat similar to the Arnoldi coefficients in this case. However, the still-unsatisfactory orthonormality and the forever-growing $b_n$'s remain problematic features in this version of the Lanczos algorithm.

\begin{figure}[t]
   \centering
\begin{subfigure}[b]{0.46\textwidth}
\centering
\includegraphics[width=\textwidth]{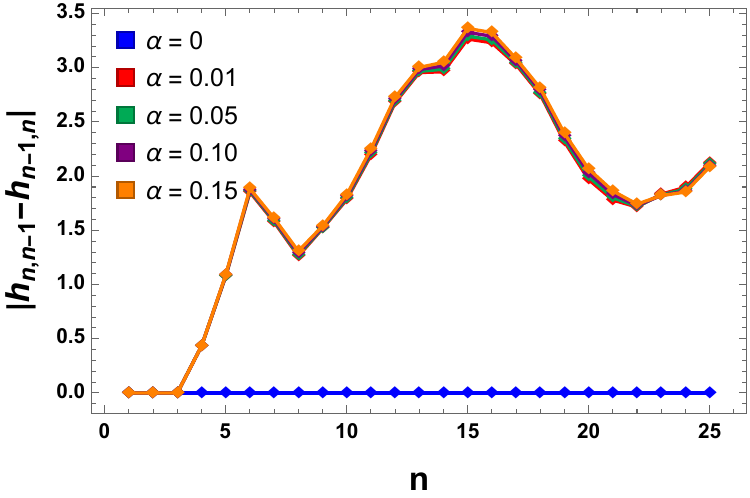}
\caption{}
\end{subfigure}
\hfill
\begin{subfigure}[b]{0.46\textwidth}
\centering
\includegraphics[width=\textwidth]{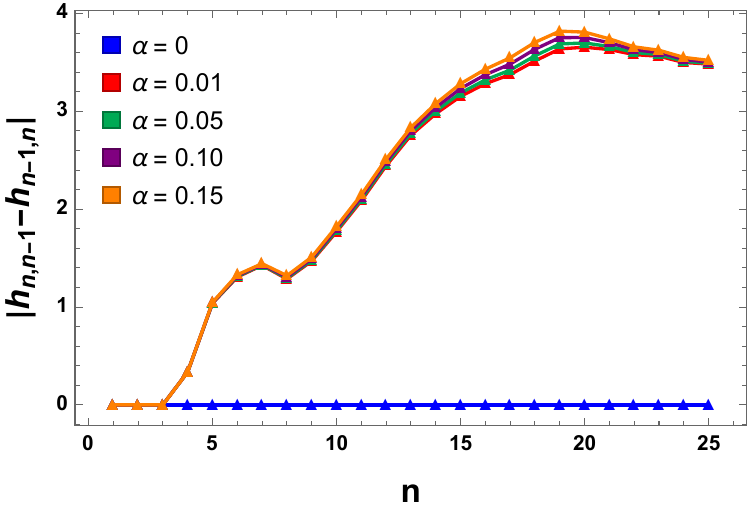}
\caption{}
\end{subfigure}
\hfill
\begin{subfigure}[b]{0.44\textwidth}
\centering
\includegraphics[width=\textwidth]{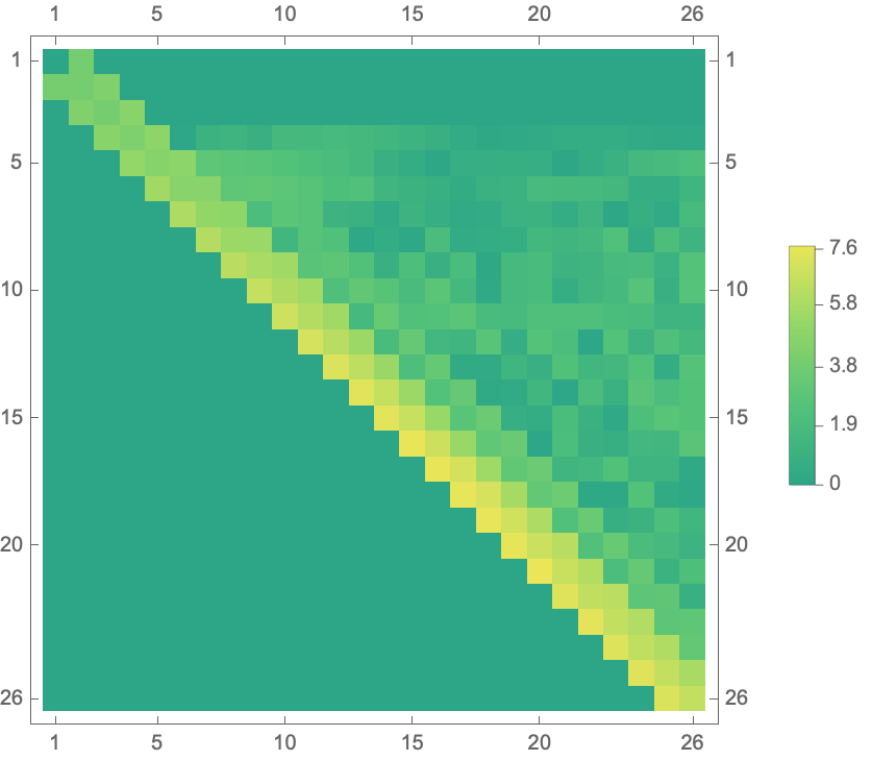}
\caption{}
\end{subfigure}
\hfill
\begin{subfigure}[b]{0.44\textwidth}
\centering
\includegraphics[width=\textwidth]{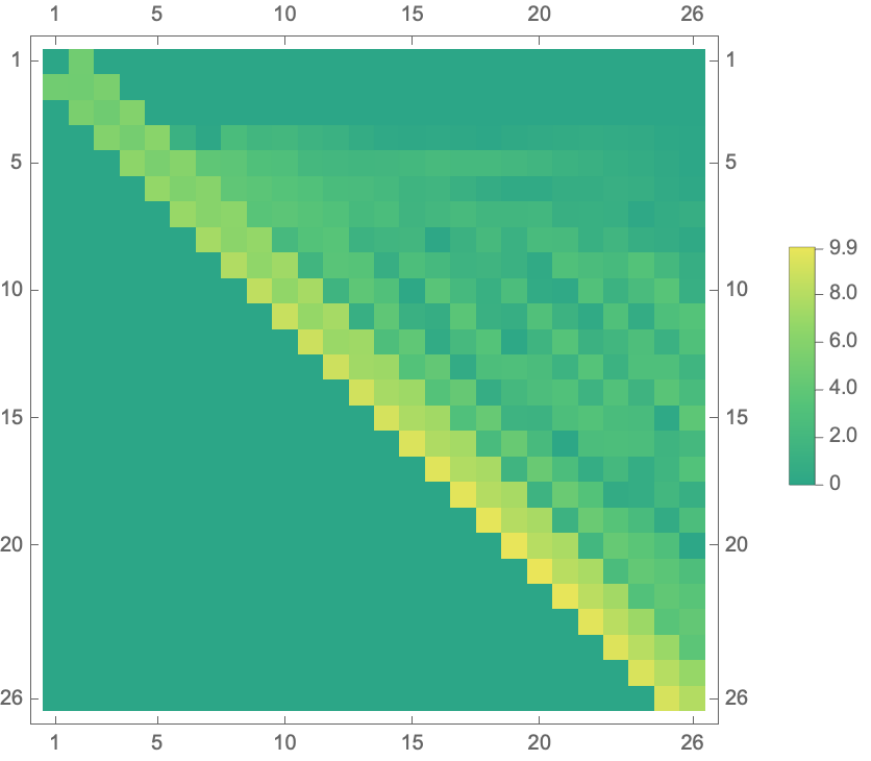}
\caption{}
\end{subfigure}
\caption{Growth of the difference between Arnoldi coefficient $|h_{n,n-1} - h_{n-1,n}|$'s of the operator $Z_3$ for (a) integrable and (b) chaotic limits for bulk dephasing $\gamma = 1$ using Arnoldi iteration. The closed system results (blue curve) corresponding to $\alpha = \gamma = 0$. We choose lattice size $N=6$. Matrix elements of the Lindbladian in (c) integrable and (d) chaotic limits.} \label{fig:high2}
\end{figure}

\subsection{Arnoldi results}
On the other hand, if we apply the Arnoldi iteration for the cases mentioned above, we find (as shown in part Fig.\,\ref{fig:Ardiagonalsi}\,(a)) that the $h_{n,n-1}$'s are almost insensitive to different kind of interaction strengths considered and always successfully distinguish integrable and chaotic regimes. Therefore the natural question that arises in one's mind is where does the information about different $\alpha$ and $\gamma$ go? As described in the main text, this is captured in the magnitudes of the purely imaginary diagonal components of the Hessenberg form. However, these diagonal components can not distinguish the integrable and chaotic regimes of the coupling $g$ and $h$ (see Fig.\,\ref{fig:Ardiagonalsi} (b)). Therefore, the information about integrability and non-Hermiticity are stored very differently through different matrix elements in the Hessenberg form.

\section{Appendix: Arnoldi iteration for high bulk dephasing} \label{appb}

In this Appendix, we briefly discuss the behavior of Arnoldi coefficients for high bulk dephasing. From Fig.\,\ref{fig:high}, we see that the coefficients $h_{n,n-1}$ clearly distinguishes the integrable and chaotic limit. They do not depend on the boundary amplitude as we previously seen. The magnitude of the coefficients also changes significantly, especially the difference between the first set of off-diagonals appears to be unaffected with increasing value of boundary amplitude damping (see Fig.\,\ref{fig:high2} (a), (b)). This appears to be true for both in integrable and chaotic limit. In Fig.\,\ref{fig:high2} (c), (d), we have shown the matrix form of the Lindbladian. We clearly see that $h_{n,n-1} \neq h_{n-1,n}$. Further, other off-diagonal elements are also non-zero, although their magnitude seems to decrease as we go far from the primary off-diagonal elements.

\section{Appendix: On the purely imaginary diagonal coefficients} \label{appc}

As mentioned previously in the main text, in the Lindbladian evolution $e^{\tilde{\mathcal{L}}_o t}$, the Lindbladian matrix $\tilde{\mathcal{L}}_o$ is known to have only real or complex conjugate pairs of eigenvalues. This representation of the Lindbladian, when compared to the evolution we studied, has the form, $\tilde{\mathcal{L}}_o=i \mathcal{L}_o$. We will therefore show here that with a tridiagonal form (once we ignore the small coefficients apart from the tridiagonalized part, which we have characterized as the spreading of the difference of $h_{n,n-1}$ and $h_{n-1,n}$ coefficients due to the artefact of the Arnoldi algorithm) of $\mathcal{L}_o$ with purely real norms $h_{n,n-1}(=h_{n-1,n}\,, \text{after ignoring the differences})$, the modified $\tilde{\mathcal{L}}_o$ can only have real and complex conjugate pairs as eigenvalues if the diagonals of the unmodified $\mathcal{L}_o$ are purely imaginary. From Eq.\,\eqref{modifiedL}, we will hereafter consider it as a simpler tridiagonal matrix of the form
\begin{equation}\label{modifiedL2}
        \tilde{\mathcal{L}_o} =\begin{pmatrix}
-|a_1| & i b_1 & 0 & \cdots & 0\\
i b_1 & -|a_2| & i b_2 & \cdots & 0\\
0 & i b_2 & -|a_3| & i b_3 & \cdots\\
\cdots & \cdots & i b_3 &\cdots &\cdots\\
0 & \cdots & \cdots &\cdots & i b_{n-1}\\
0 & 0 & \cdots & i b_{n-1}& -|a_n|\\
\end{pmatrix}\,.
    \end{equation}
With this, the characteristic equation of the above matrix turns out to be of the following form
\begin{equation}
       \mathrm{det} \begin{pmatrix}
-|a_1|-\lambda & i b_1 & 0 & \cdots & 0\\
i b_1 & -|a_2|-\lambda & i b_2 & \cdots & 0\\
0 & i b_2 & -|a_3|-\lambda & i b_3 & \cdots\\
\cdots & \cdots & i b_3 &\cdots &\cdots\\
0 & \cdots & \cdots &\cdots & i b_{n-1}\\
0 & 0 & \cdots & i b_{n-1}& -|a_n|-\lambda\\
\end{pmatrix} = 0\,,
    \end{equation}
where $\lambda$ denotes the variable in which the characteristics polynomial equation is formed, and has the eigenvalues as the roots. Now we can use the formalism of continuant \cite{continuant}, which is known to be a representation of the determinant of a tridiagonal matrix, which will be the polynomial characteristics equation of $\tilde{\mathcal{L}}_o$. Using the recurrence relation of the continuant, we can write
\begin{equation}
     K_0=1\,, ~~ \, K_1=-|a_1|-\lambda\,,~~ K_n=(-|a_n|-\lambda)K_{n-1}+b_{n-1}^2 K_{n-2}\,.
\end{equation}
This recurrence relation can be derived easily using the determinant identity of block matrices. Here the $K_n=0$ provides us with the $n$-th degree polynomial of $\lambda$ for the characteristics equation. In this equation, it is easy to see that all the coefficients are real, given that all the $|a_i|$ and $b_i$ are real. Therefore, all this equation's roots will be either real or complex conjugate pairs. At the same time, if we take $|a_i| (= |h_{i, i}|)$'s to be complex in general, this can never hold to be true. The coefficients will become complex in general, and they will not follow the complex conjugate roots theorem. According to this theorem, in a polynomial equation with all real coefficients, if there exists one complex root $p+i q$, it is guaranteed that $p-iq$ will also be a solution to the equation. This means that if we have such a polynomial equation as the following
 \begin{equation}
     f(z)=m_n z^n+ m_{n-1} z^{n-1}+\cdots + m_0=0\,,
 \end{equation}
where $m_i$ are all real, one can always show that if $z_0$ is a solution of this equation, so will be $z_0^*$ by taking a complex conjugate of the whole equation ($f(z_0)=f^*(z_0^*)$), since the real coefficients will not get modified. On the other hand, if some of the coefficients were complex, $f(z)\neq f^*(z^*)$ and therefore, it can be guaranteed that the complex conjugate pairs follow two different sets of equations and can not be roots of the same polynomial equation. In such a case, it can be shown that at least one of the complex eigenvalues will not have its conjugate pair as the eigenvalue of the polynomial equation if there is any complex coefficient. However, for the Lindbladian, the complex eigenvalues with a nonzero real part are crucial because they reflect their non-unitary nature and always come in conjugate pairs. Therefore we conclude that for the modified Lindbladian ($\tilde{\mathcal{L}}_o$) to have complex conjugate pairs of roots as eigenvalues, it is necessary that the tridiagonalized unmodified Lindbladian $\mathcal{L}_o$ has only purely imaginary diagonal entries. It is also worth noting that if we put the diagonals to zero, the Lindbladian matrix form reduces to that of a Liouvillian. All the eigenvalues are either zero or purely complex, appearing in conjugate pairs ($\pm i \theta $). This is expected since the evolution $e^{\mathcal{L}t}$, in that case, is unitary with eigenvalues $e^{\pm i \theta t}$.

\subsection{More general Hessenberg form and alternating real/imaginary entries:}

The above argument can also be extended in the presence of elements other than the tridiagonal, given that alternating diagonals are purely real and purely imaginary, as we find. To see this, one can check that if we deal with more than a tridiagonalized version, only these alternating real/ imaginary diagonals can produce a characteristics polynomial with real coefficients. If it is of any other form, it will always give rise to complex coefficients, which do not obey the complex conjugate roots theorem. In the simplest language, this is related to the fact that for this particular alternating real/imaginary diagonals form of the Hessenberg matrix $\mathcal{L}_o$, the modified Lindbladian $\tilde{\mathcal{L}}_o$ can have all the co-factors of a real entry to be real and purely imaginary entry to be purely imaginary. So, the multiplication of them always remains real. For example, we can begin with the following matrix
\begin{equation}
    \mathcal{L}_o=\left(
\begin{array}{cccccc}
 i |a_1| & b_1 & 0 & 0 & 0 & 0 \\
 c_1 & i |a_2| & b_2 & 0 & 0 & 0 \\
 0 & c_2 & i |a_3| & b_3 & i |d_1| & e_1 \\
 0 & 0 & c_3 & i |a_4| & b_4 & i |d_2| \\
 0 & 0 & 0 & c_4 & i |a_5| & b_5 \\
 0 & 0 & 0 & 0 & c_6 & i |a_6| \\
\end{array}
\right)\,,
\end{equation}
 where $h_{n,n} = i |a_{n}|$, $h_{n-1,n} = b_n$, and $h_{n,n-1}=c_n$. We have also taken other elements $i d_n$ and $e_n$ after two iterations (starting from $3$rd row) in an alternating real and imaginary way, similar to what we found in Arnoldi iteration. For this, the modified Lindbladian $\tilde{\mathcal{L}}_o=i \mathcal{L}_o$ is
 \begin{equation}
    \tilde{\mathcal{L}}_o=\left(
\begin{array}{cccccc}
 - |a_1| & ib_1 & 0 & 0 & 0 & 0 \\
 ic_1 & - |a_2| & ib_2 & 0 & 0 & 0 \\
 0 & ic_2 & - |a_3| & i b_3 & - |d_1| & ie_1 \\
 0 & 0 & ic_3 & - |a_4| & i b_4 & -|d_2| \\
 0 & 0 & 0 & ic_4 & - |a_5| & i b_5 \\
 0 & 0 & 0 & 0 & ic_6 & - |a_6| \\
\end{array}
\right)\,.
\end{equation}
For this matrix, there will be only two nontrivial co-factors we need to compute, for $(\tilde{\mathcal{L}}_o)_{11}=-|a_1|$ and $(\tilde{\mathcal{L}}_o)_{12}=i b_1$. We find the co-factors to be
\begin{align}
     \text{co-factor}[(\tilde{\mathcal{L}}_o)_{11}] = &-|a_2| (c_3 (|a_5| |a_6| b_3-c_4 (|a_6| |d_1|+c_5 e_1)+b_5 b_3 c_5) \nonumber \\
     &+|a_3| (c_4 (|a_6| b_4-c_5 d_2)+|a_4| (|a_5| |a_6| +b_5 c_5))) \nonumber \\
     &-b_2 c_2 (c_4 (|a_6| b_4-c_5 d_2)+|a_4| (|a_5| |a_6| +b_5 c_5))\,,
\end{align}    
and
\begin{align}
 \text{co-factor}[(\tilde{\mathcal{L}}_o)_{12}] = &-i c_1 (c_3 (|a_5| |a_6| b_3-c_4 (|a_6| d_1+c_5 e_1)+b_5 b_3 c_5) \nonumber \\ &+|a_3| (c_4 (|a_6| b_4-c_5 |d_2|)+|a_4| (|a_5| |a_6|+b_5 c_5)))\,.
\end{align}

 Hence, the product of a particular matrix element and its co-factor remains real in both cases. This argument can be easily extended for an arbitary dimensional matrix of this kind using mathematical induction. This realness property will not hold for any other form of the Hessenberg matrix if we change the real or imaginary nature of the alternating diagonals. Therefore, our unique form of the full Hessenberg matrix ensures that the characteristics equation has all real coefficients and eigenvalues in real or complex conjugate pairs.
 
Finally, we would like to mention that all the elements of the upper Hessenberg matrix except the primary off-diagonal elements are expected to vanish when the system size $N$ is large. This is difficult to understand in this particular example of the Ising model since, in this paper, we focussed $N=6$ only. However, in an upcoming work by one of the authors (PN) \cite{opensykupcoming}, it is indeed shown to be true in the dissipative Sachdev-Ye-Kitaev (SYK) model \cite{Kulkarni:2021gtt} by directly testing the numerical Arnoldi results with the exact analytic form. We should also mention that an effective tridiagonal form can also be obtained by implementing an oblique projection method known as the bi-Lanczos algorithm \cite{Gruning}. Here the left and right vectors are evolved separately and $h_{n,n-1}$ coefficients become equal to $h_{n-1,n}$, modulo an overall phase factor. There are no other coefficients other than the diagonal and two primary off-diagonal parts. However, this method contrasts with the Arnoldi iteration, which is known to be an orthogonal projection method. Both methods reduce to the Lanczos algorithm when the dissipative effect vanishes.

\bibliographystyle{JHEP}
\bibliography{references}

\end{document}